\documentclass[11pt,letterpaper,twoside,reqno,nosumlimits]{amsart}

\usepackage[usenames,dvipsnames]{xcolor}
\usepackage{fancyhdr}
\usepackage{amsmath,amsfonts,amsbsy,amsgen,amscd,mathrsfs,amssymb,amsthm}
\usepackage{subfig}
\usepackage{url}
\usepackage{mathtools}

\mathtoolsset{showonlyrefs}

\usepackage[font=small,margin=0.25in,labelfont={sc},labelsep={colon}]{caption}

\usepackage{tikz}
\usetikzlibrary{positioning}
\usepackage{pgfplots}
\usepackage{microtype}
\usepackage{enumitem}

\definecolor{dark-gray}{gray}{0.3}
\definecolor{dkgray}{rgb}{.4,.4,.4}
\definecolor{dkblue}{rgb}{0,0,.5}
\definecolor{medblue}{rgb}{0,0,.75}
\definecolor{rust}{rgb}{0.5,0.1,0.1}

\usepackage[colorlinks=true]{hyperref}

\hypersetup{urlcolor=rust}
\hypersetup{citecolor=dkblue}
\hypersetup{linkcolor=dkblue}

\usepackage{setspace}

\usepackage{graphicx}
\usepackage{booktabs,longtable,tabu} 
\usepackage{multirow} 

\usepackage{float}

\usepackage{fourier}
\usepackage{bm}

\graphicspath{{art/}}

\newtheorem{bigthm}{Theorem}

\newtheorem{theorem}{Theorem}[section]
\newtheorem{lemma}[theorem]{Lemma}

\newtheorem{proposition}[theorem]{Proposition}
\newtheorem{fact}[theorem]{Fact}

\newtheorem{corollary}[theorem]{Corollary}

\newtheorem*{lemma*}{Lemma}

\theoremstyle{definition}

\newtheorem{definition}[theorem]{Definition}

\newtheorem{remark}[theorem]{Remark}

\theoremstyle{remark}

\usepackage[noend]{algpseudocode}
\usepackage{algorithm,algorithmicx}

\algrenewcommand\alglinenumber[1]{\sf\scriptsize\color{blue}{#1}}
\algrenewcommand\algorithmicrequire{\textbf{Input:}}
\algrenewcommand\algorithmicensure{\textbf{Output:}}

\numberwithin{equation}{section} 
\numberwithin{figure}{section}
\numberwithin{table}{section}

\floatstyle{plaintop}
\newfloat{recipe}{thp}{lor}
\floatname{recipe}{Recipe}
\numberwithin{recipe}{section}

\providecommand{\mathbold}[1]{\bm{#1}}  
                                
\renewcommand{\phi}{\varphi}

\newcommand{\eps}{\varepsilon}

\newcommand{\Id}{\mathbf{I}}

\providecommand{\mathbbm}{\mathbb} 

\newcommand{\R}{\mathbbm{R}}
\newcommand{\C}{\mathbbm{C}}

\newcommand{\Z}{\mathbbm{Z}}
\newcommand{\Sym}{\mathbb{H}}

\newcommand{\abs}[1]{\left\vert {#1} \right\vert}

\newcommand{\Prob}[1]{\mathbbm{P}\left\{{#1}\right\}}

\newcommand{\vct}[1]{\mathbold{#1}}
\newcommand{\mtx}[1]{\mathbold{#1}}

\newcommand{\transp}{\mathsf{t}}
\newcommand{\adj}{\mathsf{t}}

\newcommand{\lspan}{\operatorname{span}}

\newcommand{\range}{\operatorname{range}}

\newcommand{\rank}{\operatorname{rank}}

\newcommand{\diag}{\operatorname{diag}}
\newcommand{\trace}{\operatorname{trace}}

\newcommand{\psdge}{\succcurlyeq}

\newcommand{\ip}[2]{\langle {#1},\ {#2} \rangle}

\newcommand{\norm}[1]{\left\Vert {#1} \right\Vert}

\newcommand{\relint}{\operatorname{relint}}

\newcommand{\conv}{\operatorname{conv}}

\newcommand{\minimize}{\mathrm{minimize}}
\newcommand{\maximize}{\mathrm{maximize}}
\newcommand{\subjto}{\mathrm{subject\ to}}

\allowdisplaybreaks

\title[Binary Component Decomposition]{Binary Component Decomposition \\ Part I: The Positive-Semidefinite Case}

\author[R.~Kueng]{Richard~Kueng}
\author[J.~A.~Tropp]{Joel~A.~Tropp}

\date{31 July 2019}

\subjclass[2010]{Primary: 52A20, 15B48. Secondary: 15A21, 52B12, 90C27.}

\keywords{Matrix factorization, cut polytope, elliptope, semidefinite programming}

\begin{document}

\begin{abstract}
This paper studies the problem of decomposing a low-rank positive-semidefinite
matrix into symmetric factors with binary entries, either $\{\pm 1\}$
or $\{0,1\}$.  This research answers fundamental questions about the existence
and uniqueness of these decompositions.  It also leads to tractable
factorization algorithms that succeed under a mild deterministic
condition. 
A companion paper addresses the related problem
of decomposing a low-rank rectangular matrix into a binary factor
and an unconstrained factor.
\end{abstract}

\maketitle

\section{Motivation}

Matrix factorization stands among the most fundamental methods for
unsupervised data analysis.  One of the main purposes of factorization
is to identify latent structure in a matrix.  Other applications include
data compression, summarization, and visualization.  In many situations,
we need to place constraints on the factors appearing in
the matrix decomposition.  This step allows us to enforce prior
knowledge about the process that generates the data, thereby enhancing
our ability to detect structure.

Prominent examples of constrained matrix factorizations include
independent component analysis~\cite{Com94:Independent-Component},
nonnegative matrix factorization~\cite{PT94:Positive-Matrix},
dictionary learning or sparse coding~\cite{OF96:Emergence-Simple-Cell},
and sparse principal component analysis~\cite{ZHT06:Sparse-PCA}.
These techniques arose in signal processing, environmental engineering, neuroscience, and statistics.
This catalog hints at the wide compass of these ideas.

In spite of the importance of constrained matrix decompositions,
researchers have only a limited understanding of which factorization
models are identifiable and which can be computed provably with efficient algorithms.
It is a natural challenge to develop rigorous theory that justifies
and improves existing factorization models.  Another valuable direction
is to create new types of constrained factorizations.
These problems not only have a deep intellectual appeal, but progress
may eventually lead to new modes of data analysis.

The purpose of this paper and its companion~\cite{KT19:Binary-Factorization-II}
is to develop the theoretical foundations for \emph{binary component decompositions}.
That is, we are interested in matrix decompositions where one of the factors is required
to take values in the set $\{ \pm 1 \}$ or in the set $\{0, 1\}$.
These models are appropriate for applications where the latent factor reflects
an exclusive choice.  For instance, ``on'' and ``off'' in electrical engineering;
``connected'' or ``disconnected'' in graph theory;
``yes'' and ``no'' in survey data;
``like'' and ``dislike'' in collaborative filtering;
or ``active'' and ``inactive'' in genomics.

We focus on core questions about the existence and uniqueness of several
types of binary factorizations, and we develop efficient
algorithms for computing these factorizations in an ideal setting.
We also report some preliminary ideas about how to obtain
binary component decompositions of noisy data.

In this first paper, we study factorization of a low-rank correlation matrix
into symmetric binary factors.  We also describe a stylized application of this decomposition
in massive MIMO communications.  In the companion paper~\cite{KT19:Binary-Factorization-II},
we build on these ideas to develop an asymmetric factorization of a low-rank data
matrix into a binary factor and an unconstrained factor.
This project takes a decisive step toward creating a new class of
matrix factorizations with a rigorous theory and implementable algorithms.

\subsection{Notation}

We use standard notation from linear algebra and optimization.
Scalars are indicated by lowercase Roman or Greek letters ($x, \xi$);
lowercase bold letters ($\vct{x}, \vct{\xi}$) are (column) vectors;
uppercase bold letters ($\mtx{X}, \mtx{\Xi}$) are matrices.
Calligraphic letters ($\mathcal{X}$) are reserved for sets.

Throughout, $n$ is a fixed natural number.
We work in the real linear space $\R^n$ equipped with the
standard inner product and the associated norm topology.
The symbol ${}^\transp$ denotes the transpose of a vector or matrix.
The standard basis in $\R^n$ is the set $\{ \mathbf{e}_1, \dots, \mathbf{e}_n \}$.
We write $\mathbf{e}$ for the vector of ones; its dimension is determined
by context.  The symbol $\odot$ denotes the Schur (i.e., componentwise) product of vectors.
The closed and open probability simplices are the sets
\begin{displaymath}
\Delta_r = \left\{ \vct{\tau} \in \R^r : \text{$\tau_i \geq 0$ and $\sum_{i=1}^r \tau_i = 1$} \right\}
\quad\text{and}\quad
\Delta_r^+ = \left\{ \vct{\tau} \in \R^r : \text{$\tau_i > 0$ and $\sum_{i=1}^r \tau_i = 1$} \right\}.
\end{displaymath}
These sets parameterize the coefficients in a convex combination.

The real linear space $\Sym_n$ consists of symmetric $n \times n$ matrices with real entries.
We write $\Id$ for the identity matrix; its dimension is determined by context.
A positive-semidefinite (psd) matrix is a symmetric matrix with nonnegative eigenvalues.
The statement $\mtx{X} \psdge \mtx{0}$ means that $\mtx{X}$ is psd.
We require an elementary property of psd matrices,
which we set down for reference.

\begin{fact}[Conjugation rule] \label{fact:psd_invariance}
Conjugation respects the semidefinite order, in the following sense.
\begin{enumerate}
\item	If $\mtx{X} \psdge \mtx{0}$, then $\mtx{KXK}^\transp \psdge \mtx{0}$ for each matrix $\mtx{K}$ with compatible dimensions.

\item	If $\mtx{K}$ has full column rank and $\mtx{KXK}^\transp \psdge \mtx{0}$, then $\mtx{X} \psdge \mtx{0}$.
\end{enumerate}
\end{fact}

\section{Sign component decomposition and binary component decomposition}

This section introduces the two matrix factorizations that we will study in
this paper, the sign component decomposition (Section~\ref{sec:scd-intro})
and the binary component decomposition (Section~\ref{sec:bcd-intro}).
It also presents our main results about situations when we can compute
these factorizations with polynomial-time algorithms.
We conclude with an outline of the paper (Section~\ref{sec:roadmap}).

\subsection{The eigenvalue decomposition}

Our point of departure is the famous eigenvalue decomposition.
Let $\mtx{A} \in \Sym_n$ be a rank-$r$ \emph{correlation matrix}.
That is, $\mtx{A}$ is a rank-$r$ psd matrix with all diagonal entries equal to one.
We can always write this matrix in the form
\begin{equation} \label{eqn:eigenvalue-decomp}
\mtx{A} = \sum_{i=1}^r \lambda_i \vct{u}_i \vct{u}_i^\transp.
\end{equation}
In this expression, 
$\{ \vct{u}_1, \dots, \vct{u}_r \} \subset \R^n$ is an orthonormal
family of eigenvectors associated with the positive eigenvalues $\lambda_1 \geq \lambda_2 \geq \dots \geq \lambda_r > 0$.
Equivalently, we may express the decomposition~\eqref{eqn:eigenvalue-decomp} as a matrix factorization:
\begin{displaymath}
\mtx{A} = \mtx{U} \, \diag(\vct{\lambda}) \, \mtx{U}^{\transp}
\quad\text{where}\quad
\mtx{U} = \begin{bmatrix} \vct{u}_1 & \dots & \vct{u}_r \end{bmatrix} \in \R^{n \times r}
\quad\text{and}\quad
\vct{\lambda} = (\lambda_1, \dots, \lambda_r). 
\end{displaymath}
The orthogonality of eigenvectors ensures that the matrix $\mtx{U}$ is {orthonormal};
that is, $\mtx{U}^\transp \mtx{U} = \Id$.

Eigenvalue decompositions are a basic tool in data analysis because
of a connection with principal component analysis.
Let $\mtx{B} \in \R^{n \times m}$ be a data matrix whose rows are standardized.%
\footnote{A vector is \emph{standardized} if its entries sum to zero and its Euclidean norm equals one.}
Then we can perform an eigenvalue decomposition~\eqref{eqn:eigenvalue-decomp}
of the correlation matrix $\mtx{A} = \mtx{BB}^\transp$ to uncover latent structure
in the columns of the data matrix $\mtx{B}$.  In this context,
the eigenvectors $\vct{u}_i$ are called \emph{principal components},
the directions in which the columns of $\mtx{B}$ vary the most~\cite{Jol02:Principal-Component}.

In spite of the significance and elegance of the decomposition~\eqref{eqn:eigenvalue-decomp},
it suffers from several debilities.
First, we cannot impose extra conditions on the eigenvectors
to enforce prior knowledge about the data.
Second, the eigenvectors are a mathematical abstraction,
so they often lack a meaning or interpretation.
Moreover, in applications, it is often hard to argue
that the data was generated from orthogonal components.
To address these limitations, we may try to develop matrix
decompositions that evince other types of structure.

\subsection{Sign component decomposition} 
\label{sec:scd-intro}

In this work, we study matrix factorization models where the
underlying components are binary-valued and need not be orthogonal.
We begin with the case where the entries of the components are
restricted to the set $\{ \pm 1 \}$.
In Section~\ref{sec:bcd-intro}, we discuss an alternative model
where the entries  are restricted to the set $\{ 0, 1 \}$.

Once again, let $\mtx{A} \in \Sym_n$ be a correlation matrix.
For some natural number $r$,
consider the problem of decomposing the correlation matrix as
a proper%
\footnote{A \emph{proper} convex combination has strictly positive coefficients.}
convex combination of rank-one sign matrices:
\begin{equation} \label{eq:binary_factorization}
\mtx{A} = \sum_{i=1}^r \tau_i \vct{s}_i \vct{s}_i^\transp
\quad\text{where}\quad
\vct{s}_i \in \{ \pm 1 \}^r
\quad\text{and}\quad
(\tau_1, \dots, \tau_r) \in \Delta_r^+.
\end{equation}
Equivalently, we may write the decomposition~\eqref{eq:binary_factorization}
as a matrix factorization:
\begin{equation} \label{eqn:scd-matrix}
\mtx{A} = \mtx{S} \, \diag(\vct{\tau}) \, \mtx{S}^\transp
\quad\text{where}\quad
\mtx{S} = \begin{bmatrix} \vct{s}_1 & \dots & \vct{s}_r \end{bmatrix} \in \{\pm 1\}^{n \times r}
\quad\text{and}\quad
\vct{\tau} = (\tau_1, \dots, \tau_r) \in \Delta_r^+.
\end{equation}
Note that the right-hand side of~\eqref{eq:binary_factorization}
always yields a correlation matrix.  
See Figure~\ref{fig:scd} for a schematic.
We refer to the factorization~\eqref{eq:binary_factorization}--\eqref{eqn:scd-matrix}
as a \emph{sign component decomposition} of the correlation matrix $\mtx{A}$.
The $\pm 1$-valued vectors $\vct{s}_i$ are called \emph{sign components},
and they may be correlated with each other.
Altogether, these properties give the factorization a combinatorial flavor,
rather than a geometric one.
See Section~\ref{sec:related-work} for a discussion of some other
discrete matrix decompositions.

\begin{figure}[t]
\begin{tikzpicture}[scale=0.6]
\draw[very thick,fill=blue!10] (1,0) rectangle (5,4);
\node at (4.5,3.5) {\Large $\mtx{A}$};
\draw[very thick] (1,4) -- (5,0);
\draw (1.5,3.5)  node[fill=blue!10] {$1$};
\draw (4.5,0.5)  node[fill=blue!10] {$1$};
\node at (6,2) {\Large $=$};
\draw[very thick,fill=blue!10] (7,0) rectangle (9,4);
\node at (8.5,3.5) {\Large $\mtx{S}$};
\node at (8,2) {\Large $\pm$};
\draw[very thick] (9.5,4) -- (11.5,2);
\draw (10,3.5) node[fill=white] {$\tau_1$};
\draw (11,2.5) node[fill=white] {$\tau_r$};
\draw[very thick] (9.5,2) rectangle (11.5,4);
\draw[very thick,fill=blue!10] (12,2) rectangle (16,4);
\node[right] at (12,3.5) {\Large $\mtx{S}^\transp$};
\node at (14,3) {\Large $\pm$};
\node at (17,2) {\Large $=$};
\node[right] at (17.5,2) {\Huge $\sum_i$}; 
\node[right] at (19.5,2)  {\LARGE $\tau_i$};
\draw[very thick,fill=blue!10] (21,0) rectangle (22,4);
\node at (21.5,3.5) {\Large $\vct{s}_i\phantom{{\vct{s}_i}^\transp\hspace{-6mm}}$};
\node at (21.5,2) {\Large $\pm$};
\draw[very thick,fill=blue!10] (22.5,3) rectangle (26.5,4);
\node[right] at (22.5,3.5) {\Large ${\vct{s}_i}^\transp$};
\node at (24.5,3.5) {\Large $\pm$};
\end{tikzpicture}
\caption{\textit{Sign component decomposition.}
The sign component decomposition~\eqref{eq:binary_factorization}--\eqref{eqn:scd-matrix}
expresses a correlation matrix $\mtx{A}$ as a proper convex combination of rank-one sign matrices.} \label{fig:scd}
\end{figure}

\subsubsection{Schur independence}

Although the sign component decomposition may appear to be combinatorially intricate,
we can compute it efficiently for a surprisingly large class of instances.  This positive
outcome stems from remarkable geometric properties of the set of correlation matrices.
Our key insight is to avoid degenerate decompositions by requiring the sign
components to be sufficiently distinct.
The following definition from~\cite{LP96:Facial-Structure} is central to our program.

\begin{definition}[Schur independence of sign vectors] \label{def:schur-independence}
A set $\left\{\vct{s}_1,\ldots,\vct{s}_r \right\} \subseteq \left\{ \pm 1 \right\}^n$ of sign vectors  
is \emph{Schur independent} if the linear hull of all pairwise Schur products has the maximal dimension:
\begin{displaymath}
\dim \lspan \big\{ \vct{s}_i \odot \vct{s}_j : 1 \leq i,j \leq r \big\}
	= \tbinom{r}{2} + 1.
\end{displaymath}
Equivalently, the family $\{ \mathbf{e} \} \cup \{ \vct{s}_i \odot \vct{s}_j : 1 \leq i < j \leq r \} \subset \R^n$
must be linearly independent.
\end{definition}

Here are some simple observations.
If a set is Schur independent, so is every subset.  Schur independence of a set 
is unaffected if we flip the sign of any subset of the vectors.
Last, it is computationally easy to check if a set of sign vectors is Schur independent.

We can interpret Definition~\ref{def:schur-independence} as a ``general position'' property for sign vectors.
A Schur independent family is always linearly independent (Lemma~\ref{lem:schur_to_linear}),
but the converse is not true in general.
Indeed, the cardinality $r$ of a Schur-independent collection of sign vectors in $\R^n$
must satisfy the bound
\begin{equation} \label{eq:factorization_rank}
r \leq \tfrac{1}{2} \big( 1 + \sqrt{ 8n - 7 } \big). 
\end{equation}
When $r$ meets the threshold~\eqref{eq:factorization_rank},
most collections of $r$ sign vectors are Schur independent.
Indeed, a randomly chosen family of sign vectors is Schur independent
with overwhelming probability.
Here is a basic result in this direction~\cite[Thm.~2.9]{Tro18:Simplicial-Faces}.

\begin{fact}[Tropp] \label{fact:schur-independence-generic}
Suppose that the vectors $\vct{s}_1, \dots, \vct{s}_r$ are drawn independently and uniformly at random from
$\{ \pm 1 \}^n$.  Then $\{ \vct{s}_1, \dots, \vct{s}_r \}$ is Schur independent with probability
at least $1 - r^2 \exp(-n/r^2)$.
\end{fact}

\noindent
See~\cite[Thms.~2.10, 2.11]{Tro18:Simplicial-Faces} for extensions to other probability models
and significant improvements.

\subsubsection{Computing a sign component decomposition}

The main outcome of this paper is an efficient algorithm for computing the
sign component decomposition of a rich class of correlation matrices.
Schur independence of the sign components is the only condition required.

\begin{bigthm}[Sign component decomposition] \label{thm:scd-main}
Let $\mtx{A} \in \Sym_n$ be a rank-$r$ correlation matrix that admits
a sign component decomposition~\eqref{eq:binary_factorization}--\eqref{eqn:scd-matrix}
where the family  $\{ \vct{s}_1, \dots, \vct{s}_r \}$ of sign components
is Schur independent.  Then the sign component decomposition is uniquely
determined up to trivial symmetries.  Algorithm~\ref{alg:symfactor}
computes the decomposition in time polynomial in $n$.
\end{bigthm}

\noindent
The uniqueness claim in Theorem~\ref{thm:scd-main} is established in Theorem~\ref{thm:simplicial_characterization}.
The computational claim is the content of Theorem~\ref{thm:correct}.

Sign component decompositions have a combinatorial quality,
and they are related to challenging combinatorial optimization problems,
such as \textsc{maxcut}~\cite{Kar72:21-NP-Problems}.
Thus, it seems surprising that this factorization is ever tractable.
Nonetheless, Theorem~\ref{thm:scd-main} asserts that we can compute
the sign component decomposition under a mild regularity condition.
This condition---Schur independence of the sign components---may seem alien at first sight,
but it is intimately related to the uniqueness of the factorization;
see Section~\ref{sub:uniqueness}. 

Theorem~\ref{thm:scd-main} only asserts that we can factorize a low-rank matrix.
We report methods for overcoming this difficulty in the companion paper~\cite{KT19:Binary-Factorization-II},
but the fundamental problem of factorizing noisy matrices remains open.
Indeed, Theorem~\ref{thm:scd-main} relies on remarkable 
geometric properties 
that are not stable under perturbation of the input matrix.
We will address this practical issue in future work.

\begin{algorithm}[t]
{\small
\begin{algorithmic}[1]
\caption{{ \small \textit{Sign component decomposition~\eqref{eq:binary_factorization} of a matrix with Schur independent components.} \newline
Implements the procedure from Section~\ref{sec:alg-intro}.}}
\label{alg:symfactor}

\Require	Rank-$r$ correlation matrix $\mtx{A}\in \Sym_n$ that satisfies~\eqref{eq:binary_factorization} with Schur independent sign components
\Ensure		Sign components $\{\tilde{\vct{s}}_1,\dots,\tilde{\vct{s}}_r\} \subseteq \{\pm 1\}^n$ and convex coefficients $\tilde{\vct{\tau}} \in \Delta_r^+$ where $\mtx{A} = \sum_{i=1}^r \tilde{\tau}_i \tilde{\vct{s}}_i \tilde{\vct{s}}_i^\transp$
\Statex
\Function{SignComponentDecomposition}{$\mtx{A}$}

\State	$[n, \sim] \gets \texttt{size}(\mtx{A})$ and $r \gets \rank(\mtx{A})$

\For{$i = 1$ to $(r-1)$}
\State	$\mtx{U} \gets \texttt{orth}(\mtx{A})$
	\Comment Find a basis for the range of $\mtx{A}$
\State	$\vct{g} \gets \texttt{randn}(n, 1)$
	\Comment Draw a random direction
\State	Find the solution $\mtx{X}_{\star}$ to the semidefinite program
\Comment	Step 1
\begin{displaymath}
\underset{\mtx{X} \in \Sym_n}{\maximize} \quad \vct{g}^\transp \mtx{X} \vct{g}
\quad\subjto\quad \text{$\trace \left( \mtx{U}^\transp \mtx{X} \mtx{U} \right) = n$, $\mathrm{diag}(\mtx{X}) = \mathbf{e}$ and
$\mtx{X} \psdge \mtx{0}$}
\end{displaymath}
\State	Factorize the rank-one matrix $\mtx{X}_{\star} = \tilde{\vct{s}}_i \tilde{\vct{s}}_i^\transp$
\Comment	Extract a sign component 
\State	Find the solution $\zeta_{\star}$ to the semidefinite program 
\Comment	Step 2
\begin{displaymath}
\underset{\zeta \in \mathbb{R}}{\maximize} \quad \zeta \quad \subjto \quad
\zeta \mtx{A} + (1-\zeta) \mtx{X}_{\star} \psdge \mtx{0}
\end{displaymath}
\State	$\mtx{A} \gets \zeta_{\star} \mtx{A} + (1-\zeta_{\star}) \mtx{X}_{\star}$
\Comment	Step 3
\EndFor
\State	Factorize the rank-one matrix $\mtx{A} = \tilde{\vct{s}}_r\tilde{\vct{s}}_r^\transp$
\Comment	$\rank(\mtx{A}) = 1$ in final iteration
\State	Find the solution $\tilde{\vct{\tau}} \in \Delta_r^+$ to the linear system
\Comment	Step 4
\begin{displaymath}
\mtx{A} = \sum_{i=1}^r \tilde{\tau}_i \tilde{\vct{s}}_i \tilde{\vct{s}}_i^\transp
\end{displaymath}
\EndFunction
\end{algorithmic}
}
\end{algorithm}

\subsection{Binary component decomposition}
\label{sec:bcd-intro}

Sign component decomposition provides a foundation for computing other types
of binary factorizations.  In particular, we can also study models where
the components take values in the set $\{0, 1\}$.  Let us summarize our
results for the latter problem.

Let $\mtx{H} \in \Sym_n$ be a psd matrix.  Our goal is to find a representation
\begin{equation} \label{eq:01_factor}
\mtx{H} = \sum_{i=1}^r \tau_i \vct{z}_i \vct{z}_i^\transp
\quad\text{where}\quad
\text{$\vct{z}_i \in \{0, 1\}^n$ and $(\tau_1, \dots, \tau_r) \in \Delta_r^+$.}
\end{equation}
Equivalently, we may write the decomposition~\eqref{eq:01_factor}
as a matrix factorization:
\begin{equation} \label{eqn:bcd-matrix}
\mtx{A} = \mtx{Z} \, \diag(\vct{\tau}) \, \mtx{Z}^\transp
\quad\text{where}\quad
\mtx{Z} = \begin{bmatrix} \vct{z}_1 & \dots & \vct{z}_r \end{bmatrix} \in \{0, 1\}^{n \times r}
\quad\text{and}\quad
\vct{\tau} = (\tau_1, \dots, \tau_r) \in \Delta_r^+.
\end{equation}
We refer to~\eqref{eq:01_factor}--\eqref{eqn:bcd-matrix}
as a \emph{binary component decomposition} of the matrix $\mtx{H}$.
The vectors $\vct{z}_i$ are called \emph{binary components}.

We can connect the binary component decomposition with the sign component decomposition
by a simple device.  Just observe that there is an affine map that places binary vectors
and sign vectors in one-to-one correspondence:
\begin{equation} \label{eqn:affine-map}
\mtx{F} : \{ 0, 1 \}^n \to \{ \pm 1 \}^n
\quad\text{where}\quad
\mtx{F} : \vct{z} \mapsto 2\vct{z} - \mathbf{e}
\quad\text{and}\quad
\mtx{F}^{-1} : \vct{s} \mapsto \tfrac{1}{2}(\vct{s} + \mathbf{e}).
\end{equation}
Owing to the correspondence~\eqref{eqn:affine-map},
Schur independence of sign vectors begets a
concept of Schur independence for binary vectors.

\begin{definition}[Schur independence of binary vectors] \label{def:schur-independence-binary}
Let $\vct{z}_0 = \mathbf{e}$.  A set $\left\{\vct{z}_1,\ldots,\vct{z}_r \right\} \subseteq \left\{ 0, 1 \right\}^n$ of binary vectors   is \emph{Schur independent} if
\begin{displaymath}
\dim \lspan \big\{ \vct{z}_i \odot \vct{z}_j : 0 \leq i,j \leq r \big\}
	= \tbinom{r}{2} + 1.
\end{displaymath}
\end{definition}

\noindent
Proposition~\ref{prop:schur-bcd-scd} describes the precise relationship between
the two notions of Schur independence.

The correspondence~\eqref{eqn:affine-map} also allows us to reduce
binary component decomposition to sign component decomposition.
The following result is a (nontrivial) corollary of Theorem~\ref{thm:scd-main}.

\begin{bigthm}[Binary component decomposition] \label{thm:bcd-main}
Let $\mtx{H} \in \Sym_n$ be a rank-$r$ psd matrix that admits
a binary component decomposition~\eqref{eq:01_factor}--\eqref{eqn:bcd-matrix}
where the family  $\{ \vct{z}_1, \dots, \vct{z}_r \}$ of binary components
is Schur independent.  Then the binary component decomposition is uniquely
determined up to trivial symmetries.  Algorithm~\ref{alg:binfactor}
computes the decomposition in time polynomial in $n$.
\end{bigthm}

\noindent
See Section~\ref{sec:bcd-compute} for the proof.

The binary component decomposition~\eqref{eq:01_factor}
is closely related to the (symmetric) cut decomposition~\cite{FK99:Cut-Decomposition,AN06:Cut-Norm}.
In general, cut decompositions seem to involve challenging
combinatorial optimization problems.  Viewed from this angle,
it seems surprising that binary component decompositions
are unique and efficiently computable.
See Section~\ref{sec:related-work} for further discussion.

It is worthwhile to point out that the regularity condition for binary components
differs slightly from its counterpart for sign components.
The vector $\mathbf{e}$ of ones features in Definition~\ref{def:schur-independence-binary}
but not in Definition~\ref{def:schur-independence}.
This modification imposes slightly more stringent conditions on binary components.
It arises from the fact that the two decompositions enjoy different symmetries:
sign vectors are invariant under flipping the global sign, while binary vectors are not.

\subsection{Roadmap}
\label{sec:roadmap}

Section~\ref{sec:exist-unique} discusses the problems of existence,
uniqueness, and computability of sign component decompositions at
a high level.  Section~\ref{sec:geometry} elaborates on the geometry
of the set of correlation matrices and its implications for sign
component decomposition.  Section~\ref{sec:scd-algorithm} proves
that Algorithm~\ref{alg:symfactor} computes a sign component
decomposition.  Section~\ref{sec:01factorization} treats
the binary component decomposition.
Afterward, in Section~\ref{sec:mimo}, we present a stylized application
to massive MIMO communication.  Section~\ref{sec:related-work}
covers related work.

\begin{algorithm}[t]
{\small
\begin{algorithmic}[1]
\caption{{\small \textit{Binary component decomposition~\eqref{eq:01_factor} of a matrix with Schur independent components}. \newline
Implements the procedure from Section~\ref{sec:bcd-compute}.}}
\label{alg:binfactor}

\Require	Rank-$r$ symmetric matrix $\mtx{H} \in \Sym_n$ that satisfies~\eqref{eq:01_factor}
with Schur independent binary components.
\Ensure		Binary components $\{ \tilde{\vct{z}}_1, \dots, \tilde{\vct{z}}_r \} \subseteq \{0,1\}^n$
and convex coefficients $\tilde{\vct{\tau}} \in \Delta_r$
where $\mtx{H} = \sum_{i=1}^r \tilde{\tau}_i \, \tilde{\vct{z}}_i\tilde{\vct{z}}_i^\transp$
\Statex
\Function{BinaryComponentDecomposition}{$\mtx{H}$}

\State	Find the solution $\mtx{A} \in \Sym_n$ to the linear system
\begin{displaymath}
\diag(\mtx{X}) = \mathbf{e}
\quad\text{and}\quad
\mtx{R} (4 \mtx{H} - \mtx{X}) \mtx{R} = \mtx{0}
\quad\text{where}\quad
\mtx{R} = \Id - n^{-1} \mathbf{ee}^\transp
\end{displaymath}

\State	Apply Algorithm~\ref{alg:symfactor} to $\mtx{A}$ to obtain
sign components $\tilde{\vct{s}}_1, \dots, \tilde{\vct{s}}_r$
and convex coefficients $\tilde{\vct{\tau}} \in \Delta_r^+$

\State	Find the solution $\vct{\xi} \in \R^r$ to the linear system
\Comment Resolve sign ambiguities
\begin{displaymath}
n \sum_{i=1}^r \tilde{\tau}_i \xi_i \tilde{\vct{s}}_i
	= (4 \mtx{H} - \mtx{A}) \mathbf{e} - 2n \trace(\mtx{H}) \mathbf{e}
\end{displaymath}

\State	Set $\tilde{\vct{z}}_i = \frac{1}{2} (\xi_i \tilde{\vct{s}}_i + \mathbf{e})$ for each index $i$
\EndFunction
\end{algorithmic}
}
\end{algorithm}

\section{Existence, uniqueness, and computation}
\label{sec:exist-unique}

This section introduces a geometric perspective on the sign component decomposition.
This view leads to our main results on existence, uniqueness, and computability.

\subsection{Questions}

We focus on three fundamental problems raised by the
definition~\eqref{eq:binary_factorization}--\eqref{eqn:scd-matrix}
of the sign component decomposition:
\begin{enumerate}
\item \textbf{Existence:} Which correlation matrices admit a sign component decomposition?
\item \textbf{Uniqueness:} When is the sign component decomposition unique, modulo symmetries?
\item \textbf{Computation:} How can we find a sign component decomposition in polynomial time? 
\end{enumerate}
The rest of this section summarizes our answers to these questions.
To make the narrative more kinetic, we postpone some standard definitions
and the details of the analysis to subsequent sections.
While the first two problems reduce to basic geometric considerations,
our investigation of the third question pilots us into more interesting territory.

There is also a fourth fundamental problem:
\begin{enumerate}[resume]
\item \textbf{Robustness:} How can we find a sign component decomposition from a noisy observation?
\end{enumerate}
We do not treat this question here,
but we present some limited results for a closely related decomposition
in the companion work~\cite{KT19:Binary-Factorization-II}.
Understanding robustness is a critical topic for future research.

\subsection{Existence of the sign component decomposition} 

The first order of business is to delineate circumstances in which
a correlation matrix admits a sign component decomposition.

To that end, we introduce the \emph{elliptope}, the set
of all correlation matrices with fixed dimension:
\begin{displaymath}
\mathcal{E}_n = \left\{ \mtx{X} \in \Sym_n : \text{$\mathrm{diag}(\mtx{X}) = \mathbf{e}$ and $\mtx{X} \psdge \mtx{0}$} \right\}.
\end{displaymath}
The geometry of the elliptope plays a central role in our development,
so we take note of some basic properties.
The elliptope $\mathcal{E}_n$ is a compact convex subset of $\Sym_n$,
and we can optimize a linear functional over the elliptope using a
simple semidefinite program.
Among other things, the elliptope $\mathcal{E}_n$ contains each rank-one sign matrix
$\vct{ss}^\transp$ generated by a sign vector $\vct{s} \in \{\pm 1\}^n$.
In fact, each rank-one sign matrix is an extreme point of the elliptope.

Next, let us construct the set of correlation matrices
that admit a sign component decomposition.
The (signed) \emph{cut polytope} is the convex hull of the rank-one sign matrices:
\begin{equation} \label{eqn:cut-polytope}
\mathcal{C}_n = \mathrm{conv} \left\{ \vct{s} \vct{s}^\transp :  \vct{s} \in \left\{ \pm 1 \right\}^n \right\}
	\subset \Sym_n.
\end{equation}
It is easy to verify that the extreme points of the cut polytope are precisely
the rank-one sign matrices.
Since each rank-one sign matrix belongs to the elliptope, convexity ensures
that the cut polytope is contained in the elliptope: $\mathcal{C}_n \subset \mathcal{E}_n$.
This inclusion is strict.  In view of these relationships, we can think about
the elliptope as a semidefinite relaxation of the cut polytope.

The next statement is an immediate consequence
of~\eqref{eq:binary_factorization} and~\eqref{eqn:cut-polytope}.

\begin{proposition}[Sign component decomposition: Existence] \label{prop:scd-exist}
A correlation matrix $\mtx{A} \in \mathcal{E}_n$ admits
a sign component decomposition~\eqref{eq:binary_factorization}
if and only if $\mtx{A} \in \mathcal{C}_n$.
\end{proposition}

\noindent
This simple result masks the true difficulty of the problem
because the cut polytope is a very complicated object.
In fact, it is computationally hard just to decide whether
a given correlation matrix belongs to the cut polytope~\cite{DL97:Geometry-Cuts}.

\subsection{Symmetries of the sign component decomposition}

Proposition~\ref{prop:scd-exist} tells us that each matrix in the cut polytope
admits a sign component decomposition.  The next challenge is to
understand when this decomposition is determined uniquely.

First, observe that each sign component decomposition $\mtx{A} = \sum_{i=1}^r \tau_i \vct{s}_i \vct{s}_i^\transp$
has a parametric representation $(\tau_i, \vct{s}_i)$ for $i = 1, \dots, r$.  In this representation,
$r$ is a natural number, $(\tau_1, \dots, \tau_r) \in \Delta_r^+$,
and the sign components $\vct{s}_i \in \{ \pm 1 \}^n$.
But there is no way to distinguish an ordering of the pairs
($i \mapsto \pi(i)$ for a permutation $\pi$) or to distinguish
the global sign of a sign component ($\vct{s}_i \mapsto \xi_i \vct{s}_i$ for $\xi_i \in \{\pm1\}$).
Therefore, we regard two parametric representations as \emph{equivalent}
if they have the same number of terms and the terms coincide up to permutations
and sign flips.

In summary, a correlation matrix has a \emph{unique} sign component decomposition
if the parametric representation of every possible sign component decomposition
belongs to the same equivalence class.

\subsection{Uniqueness of the sign component decomposition} \label{sub:uniqueness} 

Geometrically, the sign component decomposition~\eqref{eq:binary_factorization}
is a representation of a matrix $\mtx{A} \in \mathcal{C}_n$ as a proper convex combination
of the extreme points of the cut polytope, namely the rank-one sign matrices.
The representation is unique if and only if the participating extreme points
generate a simplicial face of the cut polytope.

\begin{proposition}[Sign component decomposition: Uniqueness] \label{prop:uniqueness}
A matrix $\mtx{A} \in \mathcal{C}_n$ admits a unique
sign component decomposition~\eqref{eq:binary_factorization}
if and only if $\mtx{A}$ belongs to the relative interior
of a simplicial face of the cut polytope $\mathcal{C}_n$.
\end{proposition}

\noindent
See Section~\ref{sec:simplices} for the definition of a simplicial face;
Proposition~\ref{prop:uniqueness} follows from the discussion there.

Unfortunately, there is no simple or computationally tractable description of the simplicial
faces of the cut polytope~\cite{DL97:Geometry-Cuts}.  As a consequence, we cannot expect
to produce a sign component decomposition of a general element of the cut polytope,
even when the decomposition is uniquely determined.

Instead, let us focus on simplicial faces of the elliptope that are generated by rank-one sign matrices.
These distinguished faces are always simplicial faces of the cut polytope 
because $\mathcal{C}_n \subset \mathcal{E}_n$ and the rank-one sign matrices are extreme points
of both sets.  Thus, Proposition~\ref{prop:uniqueness} has the following consequence.

\begin{corollary}[Sign component decomposition: Sufficient condition for uniqueness] \label{cor:uniqueness}
For a family $\mathcal{S} = \{ \vct{s}_1, \dots, \vct{s}_r \} \subseteq \{ \pm 1 \}^n$ of sign vectors,
suppose that $\mathcal{F} = \conv \{ \vct{ss}^\transp : \vct{s} \in \mathcal{S} \}$
is a simplicial face of the elliptope $\mathcal{E}_n$.
If $\mtx{A}$ belongs to the relative interior of $\mathcal{F}$,
then $\mtx{A}$ admits a unique sign component decomposition~\eqref{eq:binary_factorization}.
\end{corollary}

\noindent
See Section~\ref{sec:simplicial-faces} for further details.

\subsection{Simplicial faces of the elliptope}

This is where things get interesting.
Corollary~\ref{cor:uniqueness} suggests that we shift our attention
to those correlation matrices that belong to a simplicial face of the
elliptope that is generated by rank-one sign matrices.
This class of matrices admits a beautiful characterization.

\begin{theorem}[Simplicial faces of the elliptope: Characterization] \label{thm:simplicial_characterization}
Let $\mathcal{S} = \{ \vct{s}_1,\ldots,\vct{s}_r \} \subseteq \left\{ \pm 1 \right\}^n$ be a set of sign vectors.
The following are equivalent:
\begin{enumerate}
\item The family $\mathcal{S}$ of sign vectors is Schur independent.
\item The set $ \mathcal{F} = \conv \{ \vct{ss}^\transp : \vct{s} \in \mathcal{S} \}$
is a simplicial face of the elliptope $\mathcal{E}_n$.
\end{enumerate}
Either condition implies that each correlation matrix in the relative interior
of $\mathcal{F}$ has a unique sign component decomposition~\eqref{eq:binary_factorization}.
\end{theorem}

\noindent
The implication $(1) \Rightarrow (2)$ was established by Laurent and Poljak \cite{LP96:Facial-Structure};
The reverse direction $(2) \Rightarrow (1)$ is new; see Section~\ref{sec:simplicial-faces} for the proof.
The last statement is the content of Corollary~\ref{cor:uniqueness}.

To summarize, when $r$ satisfies~\eqref{eq:factorization_rank}, almost all families of $r$ sign vectors in $\R^n$
are Schur independent.  The convex hull of the associated rank-one sign matrices forms a simplicial face of the
elliptope.  Every correlation matrix in the relative interior of this face admits a unique sign component decomposition.
The problem is how to find the decomposition.

\begin{remark}[Other kinds of simplicial faces]
The elliptope has simplicial faces that are not described by Theorem~\ref{thm:simplicial_characterization}.
Indeed, for $n \geq 5$, the elliptope $\mathcal{E}_n$ has edges that are
not generated as the convex hull of two rank-one sign matrices;
see~\cite[Example~3.3]{LP96:Facial-Structure}.
\end{remark}

\subsection{Separating simplicial faces from the elliptope}

As we have seen, the elliptope has an enormous number of simplicial faces that are generated
by rank-one sign matrices.  Remarkably, we can produce an explicit
linear functional that exposes this type of face.  This construction allows
us to optimize over these distinguished simplicial faces, 
which is the core ingredient in our algorithm for sign component decomposition.

\begin{theorem}[Simplicial faces of the elliptope: Finding a separator] \label{thm:implicit}
Fix a Schur independent family $\mathcal{S} = \{ \vct{s}_1, \dots, \vct{s}_r \} \subseteq \{ \pm 1 \}^n$
of sign vectors, and let $\mtx{P} \in \Sym_n$ be the orthogonal projector onto
$\lspan \mathcal{S}$.
Construct the linear functional
\begin{displaymath}
\psi(\mtx{X}) = n^{-1} \trace(\mtx{PX})
\quad\text{for $\mtx{X} \in \Sym_n$.}
\end{displaymath}
Then $\psi$ exposes the simplicial face
$\mathcal{F} = \conv \{ \vct{ss}^\transp : \vct{s} \in \mathcal{S} \}$
of the elliptope $\mathcal{E}_n$.  That is,
\begin{displaymath}
\text{$\psi (\mtx{X}) \leq 1$ for all $\mtx{X} \in \mathcal{E}_n$}
\quad \text{and} \quad
\mathcal{F} = \{ \mtx{X} \in \mathcal{E}_n : \psi (\mtx{X}) = 1 \}.
\end{displaymath}
\end{theorem}

\begin{figure}
\begin{center}
\includegraphics[width=0.6\textwidth]{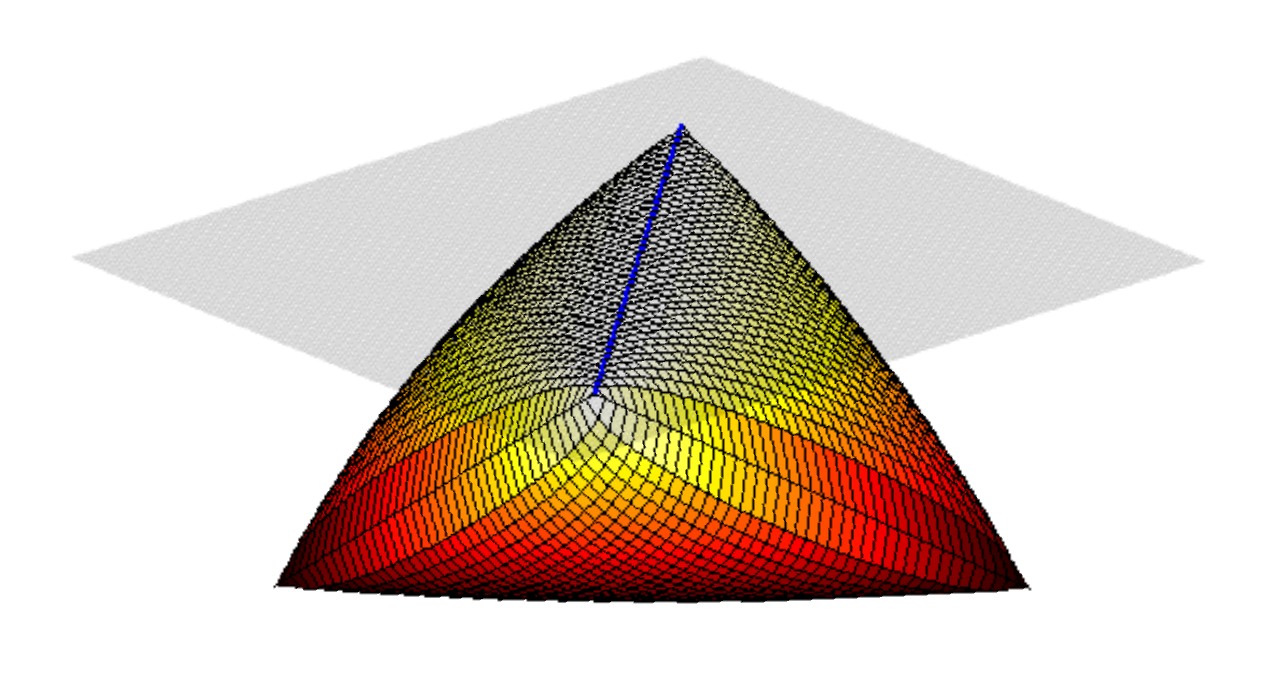}
\end{center}
\caption{\textit{Exposing a simplicial face of the elliptope.}
The hyperplane [\textit{gray}] separates a one-dimensional simplicial face
[\textit{blue}] from the elliptope $\mathcal{E}_3$
[\textit{orange}].}
\label{fig:elliptope_tangent}
\end{figure}

\noindent
See Section~\ref{sec:separator} for the proof.  See Figure~\ref{fig:elliptope_tangent}
for an illustration.

\subsection{Computing the sign component decomposition}
\label{sec:alg-intro}

With this preparation, we may now present an algorithm that computes
the sign component decomposition~\eqref{eq:binary_factorization} of a correlation matrix whose
sign components are Schur independent.  The procedure is iterative,
and it can be regarded as an algorithmic implementation
of Carath{\'e}odory's theorem~\cite[Thm.~1.1.4]{Sch13:Convex-Bodies} or 
a variant of the Gr{\"o}tschel--Lov{\'a}sz--Schrijver
method~\cite[Thm.~6.5.11]{GLS93:Combinatorial-Optimization}.

\begin{figure}
\begin{center}
\begin{tabular}{cc}
\begin{tikzpicture}[baseline,scale=0.7]
\begin{scope}[rotate=70]
\draw[-latex, very thick,color=gray] (2.5,-0.3) -- (2.5,6.5);
 \foreach \y in {0,0.5,1,...,6}
	\draw[line width=0.1 mm,color=gray,opacity=0.5] (-3,\y) -- (3,\y);
\node[left] at (2.5,6.5) {\textcolor{gray}{$\vct{g}\vct{g}^\transp$}};
\end{scope}
\begin{scope}[shift={(-3.15,1.6)},rotate = 30]
\draw[very thick] (-2,-2) -- (2,-3) -- (0,2) -- (-2,-2);
\fill[color=blue,opacity=0.1] (-2,-2) -- (2,-3) -- (0,2) -- (-2,-2);
\draw[fill=black] (-2,-2) circle (2pt);
\node at (-2.2,-2.3) {$\vct{s}_1 {\vct{s}_1}^\transp$};
\draw[fill=black] (2,-3) circle (2pt);
\node at (2,-3.4) {$\vct{s}_3 {\vct{s}_3}^\transp$};
\draw[fill=red, thick] (0,2) circle (2pt);
\node[below left] at (0,2.4) {\textcolor{red}{$\vct{s}_2 {\vct{s}_2}^\transp$}};
\draw[fill=black] (0,0) circle (2pt);
\node at (0.3,-0.2) {\Large $\mtx{A}$};
\node[blue!50!black] at (1.25, -2.3) {\Large $\mathcal{F}$};
\end{scope}
\end{tikzpicture}
 &
\begin{tikzpicture}[baseline,scale=0.7]
\begin{scope}[shift={(-2,1.4)},rotate = 30]
\draw[very thick] (-2,-2) -- (2,-3) -- (0,2) -- (-2,-2);
\fill[color=blue,opacity=0.1] (-2,-2) -- (2,-3) -- (0,2) -- (-2,-2);
\draw[fill=black] (-2,-2) circle (2pt);
\node at (-2.2,-2.3) {$\vct{s}_1 {\vct{s}_1}^\transp$};
\draw[fill=black] (2,-3) circle (2pt);
\node at (2,-3.4) {$\vct{s}_3 {\vct{s}_3}^\transp$};
\draw[-latex, very thick, color=red] (0,2) -- (0,-2.45);
\draw[fill=red, thick] (0,2) circle (2pt);
\node[below left] at (0,2.4) {\textcolor{red}{$\vct{s}_2 {\vct{s}_2}^\transp$}};
\draw[fill=red, thick] (0,0) circle (2pt);
\node at (0.3,-0.2) {\Large \textcolor{black}{$\mtx{A}$}};
\draw[fill=red, thick] (0,-2.5) circle (2pt);
\node[below] at (0,-2.8) {\Large \textcolor{red}{$\mtx{A}'$}};
\node[blue!50!black] at (1.25, -2.3) {\Large $\mathcal{F}$};
\end{scope}
\end{tikzpicture} \\
\\
\end{tabular}
\end{center}
\caption{\textit{Illustration of Algorithm~\ref{alg:symfactor}.}
The matrix $\mtx{A}$ belongs to the relative interior of the simplex $\mathcal{F}=\mathrm{conv} \left\{ \vct{s}_1 \vct{s}_1^\transp,\vct{s}_2 \vct{s}_2^\transp,\vct{s}_3 \vct{s}_3^\transp \right\}$.
[\textit{left}] Random optimization over the simplex $\mathcal{F}$ identifies an extreme point with probability one.  In this diagram, maximizing the linear functional $\mtx{X} \mapsto \trace(\vct{gg}^\transp \mtx{X})$ over $\mathcal{F}$
locates the rank-one matrix $\vct{s}_2 \vct{s}_2^\transp$.
[\textit{right}] To remove the contribution of the rank-one matrix $\vct{s}_2 \vct{s}_2^\transp$ from the matrix $\mtx{A}$, we traverse the ray from the rank-one matrix through the matrix $\mtx{A}$ until we arrive at a facet of $\mathcal{F}$.  The terminus $\mtx{A}'$ of the ray is a proper convex combination of the remaining rank-one matrices.}
\label{fig:caricature}
\end{figure}

Assume that we are given a correlation matrix $\mtx{A} \in \mathcal{E}_n$
that admits a sign component decomposition 
with Schur independent sign components:
\begin{equation} \label{eqn:A-for-alg}
\mtx{A} = \sum_{i=1}^r \tau_i \vct{s}_i \vct{s}_i^\transp
\quad\text{where}\quad
\text{$\mathcal{S} = \{ \vct{s}_1, \dots, \vct{s}_r \} \subseteq \{ \pm 1 \}^n$
is Schur independent.}
\end{equation}
As usual, the coefficients $(\tau_1, \dots, \tau_r) \in \Delta_r^+$.
The matrix $\mtx{A}$ belongs to the relative interior of the set
$\mathcal{F} = \conv\{\vct{ss}^\transp : \vct{s} \in \mathcal{S}\}$.
Theorem~\ref{thm:simplicial_characterization} implies that
$\mathcal{F}$ 
is a simplicial face of the elliptope $\mathcal{E}_n$
and the sign component decomposition of $\mtx{A}$ is unique.
Theorem~\ref{thm:implicit} allows us to formulate optimization
problems over the set $\mathcal{F}$.

The following procedure exploits these insights to identify
the sign component decomposition of $\mtx{A}$.
Figure~\ref{fig:caricature} illustrates the geometry,
while Algorithm~\ref{alg:symfactor} provides pseudocode.

\vspace{0.5pc}

\begin{itemize} \setlength{\itemsep}{0.5pc}
\item	\textbf{Step 0: Initialization.}  Let $\mtx{A} \in \mathcal{E}_n$ be a rank-$r$ correlation
matrix of the form~\eqref{eqn:A-for-alg}.  Compute the orthogonal projector $\mtx{P} \in \Sym_n$
onto $\range(\mtx{A}) = \lspan \mathcal{S}$, and set $\psi(\mtx{X}) = n^{-1} \trace(\mtx{PX})$.

\item \textbf{Step 1: Random optimization.}
Draw a (standard normal) random vector $\vct{g} \in \R^n$.
Find a solution to the semidefinite program
\begin{equation}  \label{eq:random_optimization}
\underset{\mtx{X} \in \Sym_{n}}{\maximize}  \quad \vct{g}^\transp \mtx{X}\vct{g} \quad
\subjto  \quad \text{$\psi(\mtx{X})=1$ and $\mtx{X} \in \mathcal{E}_n$.}
\end{equation}
According to Theorem~\ref{thm:implicit}, the constraint set is precisely the simplex $\mathcal{F}$.
With probability one, the unique solution $\mtx{X}_{\star}$ is an extreme point of $\mathcal{F}$.
That is, $\mtx{X}_\star = \vct{s}_k \vct{s}_k^\transp$ for some index $1 \leq k \leq r$.
By factorizing $\mtx{X}_{\star}$, we can extract one sign component of the matrix $\mtx{A}$.

\item \textbf{Step 2: Deflation.} Draw a ray from the identified factor $\mtx{X}_{\star}$ through the matrix $\mtx{A}$. Traverse this ray until we reach a facet $\mathcal{F}'$ of the simplex $\mathcal{F}$
by finding the solution $\zeta_{\star}$ of
\begin{equation} \label{eq:deflation}
\underset{\zeta \in \R}{\maximize} \quad \zeta \quad \subjto \quad \zeta \mtx{A} + (1-\zeta) \mtx{X}_\star \in \mathcal{F}.
\end{equation}
In our context, this optimization problem can be simplified,
as stated in Algorithm~\ref{alg:symfactor}.

\item	\textbf{Step 3: Iteration.} Let $\mtx{A}' = \zeta_\star \mtx{A} + (1-\zeta_\star) \mtx{X}_\star$
be the terminus of the ray described in the last step.  This construction ensures that $\mtx{A}'$ belongs to the relative
interior of the convex hull of all the rank-one sign matrices other than $\mtx{X}_{\star}$.  That is,
\begin{displaymath}
\mtx{A}' \in \relint \conv\big\{ \vct{s}_i \vct{s}_i^\transp : \text{$1 \leq i \leq r$ and $i \neq k$} \big\} = \mathcal{F}'.
\end{displaymath}
Therefore, $\mtx{A}'$ admits a sign component decomposition with Schur independent sign components.
We may return to Step 0 and repeat the process with the rank-$(r-1)$ correlation matrix $\mtx{A}'$.
The total number of iterations is $r$.

\item	\textbf{Step 4: Coefficients.} Given the $r$ computed sign components
$\tilde{\vct{s}}_1, \dots, \tilde{\vct{s}}_r$,
we can identify the convex coefficients $\tilde{\vct{\tau}} \in \Delta_r^+$ by finding the unique solution
to the linear system
\begin{displaymath}
\mtx{A} = \sum_{i=1}^r \tilde{\tau}_i \tilde{\vct{s}}_i \tilde{\vct{s}}_i^\transp.
\end{displaymath}
\end{itemize}

The following theorem states that this procedure yields a parametric representation
of the unique sign component decomposition of the matrix $\mtx{A}$.

\begin{theorem}[Analysis of Algorithm~\ref{alg:symfactor}] \label{thm:correct}
Let $\mtx{A} \in \mathcal{E}_n$ be a correlation matrix that admits
a sign component decomposition
\begin{equation} \label{eq:symmetric_factorization_restatement}
\mtx{A} = \sum_{i=1}^r \tau_i \vct{s}_i \vct{s}_i^\transp
\quad\text{where}\quad
\text{$\vct{s}_i \in \{ \pm 1 \}^n$ and $(\tau_1, \dots, \tau_r) \in \Delta_r^+$.}
\end{equation}
Assume that the family $\mathcal{S} = \{ \vct{s}_1, \dots, \vct{s}_r \}$
of sign components is Schur independent.  Then, with probability one, Algorithm~\ref{alg:symfactor}
identifies the sign component decomposition of $\mtx{A}$ up to trivial symmetries.
That is, the output is an unordered set of pairs
$ \{ (\tau_i, \xi_i \vct{s}_i ) : 1 \leq i \leq r \}$,
where $\xi_i \in \{ \pm 1 \}$ are signs.
\end{theorem}

Section~\ref{sec:scd-algorithm} contains a full proof of this result.

\begin{remark}[Certificate of uniqueness]
Given a sign component decomposition of a correlation matrix,
it is straightforward to check whether the sign components compose
a Schur independent family.  As a consequence, we can use
Theorem~\ref{thm:simplicial_characterization} to confirm
\emph{a posteriori} that we have obtained the unique
sign component decomposition of the matrix.
\end{remark}

\section{Geometric aspects of the sign component decomposition}
\label{sec:geometry}

This section contains a rigorous justification of
the geometric claims propounded in the last section.
The books~\cite{Roc70:Convex-Analysis,HL01:Fundamentals-Convex,
Bar02:Course-Convexity,Gru07:Convex-Geometry,Sch13:Convex-Bodies}
serve as good references for convex geometry.

\subsection{Faces of convex sets}

In this section, we work in a finite-dimensional real vector space $\mathsf{V}$,
equipped with a norm topology.  Let us begin with some basic facts about the boundary
structure of a convex set.

\begin{definition}[Face] \label{def:face}
Let $\mathcal{K}$ be a closed convex set in $\mathsf{V}$. A \emph{face} $\mathcal{F}$ of $\mathcal{K}$
is a convex subset of $\mathcal{K}$ for which
\begin{displaymath}
\vct{x},\vct{y} \in \mathcal{K} \quad \textrm{and} \quad \tau \vct{x} + (1-\tau) \vct{y} \in \mathcal{F} \textrm{ for some $\tau \in (0,1)$} \quad  \textrm{imply} \quad \vct{x},\vct{y} \in \mathcal{F}.
\end{displaymath}
In words, an average of points in $\mathcal{K}$ belongs to $\mathcal{F}$
if and only if the points themselves belong to $\mathcal{F}$.
\end{definition}

The faces of a closed convex set $\mathcal{K}$ are again closed convex sets.
The 0-dimensional faces are commonly called \emph{extreme points},
and 1-dimensional faces are \emph{edges}.
The set $\mathcal{K}$ is a face of itself with maximal dimension,
while faces of $\mathcal{K}$ with one lower dimension are called \emph{facets}.

Faces have a number of important properties. 
From the definition, it is clear that the ``face of'' relation is transitive: if $\mathcal{F}'$ is a face of $\mathcal{F}$ and $\mathcal{F}$ is a face of $\mathcal{K}$,
then $\mathcal{F}'$ is a face of $\mathcal{K}$.
The next fact states that the faces of a closed convex set partition the set;
see~\cite[Thm.~2.12]{Sch13:Convex-Bodies} for the proof.

\begin{fact}[Facial decomposition] \label{fact:face-decomp}
Let $\mathcal{K} \subseteq \mathsf{V}$ be a closed convex set.
Every point in $\mathcal{K}$ is contained in the relative interior
of a unique face of $\mathcal{K}$.
\end{fact}

We will also need to consider a special type of face.

\begin{definition}[Exposed face] \label{def:exposed_face}
Let $\mathcal{K}$ be a closed convex set in $\mathsf{V}$. A subset $\mathcal{F} \subseteq \mathcal{K}$
is called an \emph{exposed face} of $\mathcal{K}$ if there is a linear functional
$\psi: \mathsf{V} \to \R$ such that $\mathcal{F} = \{ \vct{x} \in \mathcal{K} : \psi (\vct{x})=1 \}$.
\end{definition}

\noindent
Exposed faces of $\mathcal{K}$ are always faces of $\mathcal{K}$,
but the converse is not true in general.

\subsection{Simplices}
\label{sec:simplices}

A \emph{simplex} is the convex hull of an affinely independent point set.
We frequently refer to \emph{simplicial faces}
of a convex set, by which we mean faces of the set that are also simplices.
The following result gives a complete description
of the faces of a simplex; see~\cite[Chap.~VI.1]{Bar02:Course-Convexity}.

\begin{fact}[Faces of a simplex] \label{fact:face-simplex}
Let $\mathcal{P} = \conv \{ \vct{x}_1, \dots, \vct{x}_N \}$ be a simplex in $\mathsf{V}$.
For each subset $I \subseteq \{1, \dots, N\}$, the set $\conv \{ \vct{x}_i : i \in I \}$ is a
simplicial face of $\mathcal{P}$.  Moreover, every face of $\mathcal{P}$ takes this form.
\end{fact}

A related result holds for the simplicial faces of more general convex sets.

\begin{lemma} \label{lem:simplicial_faces_hereditary}
Suppose that $\mathcal{F} \subseteq \mathcal{K}$ is a simplicial face of a closed convex set $\mathcal{K}$.
Then every face of $\mathcal{F}$ must also be a simplicial face of $\mathcal{K}$.
\end{lemma}

\begin{proof}
By transitivity, a
face $\mathcal{F}'$ of $\mathcal{F}$ is also a face of $\mathcal{K}$.
By Fact~\ref{fact:face-simplex}, $\mathcal{F}'$ is a simplex.
\end{proof}

\subsection{Uniqueness of convex decompositions}

Simplices are intimately related to the uniqueness of convex decompositions.
Together, Minkowski's theorem~\cite[Cor.~1.4.5]{Sch13:Convex-Bodies} and
Carath{\'e}odory's theorem \cite[Thm.~1.1.4]{Sch13:Convex-Bodies}
ensure that every point in 
a compact convex set can be written as a proper
convex combination of an affinely independent family of extreme points.
Each of these representations is uniquely determined (up to the ordering of the extreme points)
if and only if the set is a simplex.

\begin{lemma}[Unique decomposition of all points] \label{lem:uniqueness_simplicial}
Let $\mathcal{K} \subset \mathsf{V}$ be a compact convex set.
Each one of the points in the relative interior of $\mathcal{K}$ enjoys a unique
decomposition as a proper convex combination of extreme points of $\mathcal{K}$
if and only if $\mathcal{K}$ is a simplex.
\end{lemma}

\begin{proof}
Assume that $\mathcal{K}$ is a simplex.  Then $\mathcal{K} = \conv \mathcal{X}$,
where $\mathcal{X} = \{ \vct{x}_1, \dots, \vct{x}_N \} \subset \mathsf{V}$ is an affinely
independent family.  Using the definition of an extreme point, it is
easy to verify that the extreme points of $\mathcal{K}$ are precisely
the elements of $\mathcal{X}$.  Now, for any point $\vct{y}$ in the affine hull
of $\mathcal{X}$, we can find a representation of $\vct{y}$ as an affine
combination of the points in $\mathcal{X}$ by solving the linear system
\begin{displaymath}
\sum_{i=1}^N \alpha_i \vct{x}_i = \vct{y}
\quad\text{and}\quad
\sum_{i=1}^N \alpha_i = 1.
\end{displaymath}
Since the family $\mathcal{X}$ is affinely independent, this linear system is nonsingular,
and its solution is uniquely determined.  By~\cite[Lem.~1.1.12]{Sch13:Convex-Bodies},
the representing coefficients $\alpha_1, \dots, \alpha_N$
are positive precisely when $\vct{y}$ belongs to the relative interior
of the simplex $\mathcal{K} = \conv \mathcal{X}$.

For the converse, assume that $\mathcal{K}$ is not a simplex.
By Minkowski's theorem~\cite[Thm.~1.4.5]{Sch13:Convex-Bodies},
we can express $\mathcal{K} = \conv \mathcal{X}$,
where $\mathcal{X}$ is the set of extreme points of $\mathcal{K}$.
Since $\mathcal{K}$ is not a simplex, $\mathcal{X}$ is not
affinely independent.
Radon's theorem~\cite[Thm.~1.1.5]{Sch13:Convex-Bodies}
ensures that there are two finite, disjoint subsets 
of $\mathcal{X}$ whose convex hulls intersect.  Each point
in the intersection lacks a unique representation as
a proper convex combination of extreme points of $\mathcal{K}$.
\end{proof}

We can now give a precise description
of when a specific point in a polytope admits a unique decomposition.  This
following result is a direct consequence of 
Fact~\ref{fact:face-decomp} and Lemma~\ref{lem:uniqueness_simplicial}.

\begin{proposition}[Unique decomposition of one point] \label{prop:polytope-unique}
Let $\mathcal{K} \subset \mathsf{V}$ be a compact convex set, and fix a point
$\vct{x} \in \mathcal{K}$. Then the point $\vct{x}$
admits a unique decomposition as a proper convex combination of extreme points of $\mathcal{K}$
if and only the point $\vct{x}$ is contained in the relative interior of a
simplicial face of $\mathcal{K}$. 
\end{proposition}

\begin{proof}[Proposition~\ref{prop:polytope-unique}]
Let $\vct{x} \in \mathcal{K}$.  According to Fact~\ref{fact:face-decomp},
the point $\vct{x}$ belongs to the relative interior of a unique face $\mathcal{F}$
of $\mathcal{K}$.  By Definition~\ref{def:face} of a face, the point $\vct{x}$
can be written as a proper convex combination of extreme points in $\mathcal{K}$ if and only if
the participating extreme points all belong to $\mathcal{F}$.
Lemma~\ref{lem:uniqueness_simplicial} promises that $\vct{x}$ has a unique representation
as a proper convex combination of the extreme points of $\mathcal{F}$ if and only if $\mathcal{F}$
is a simplex.
\end{proof}  

Proposition~\ref{prop:uniqueness} is just the specialization
of Proposition~\ref{prop:polytope-unique} to the cut polytope $\mathcal{C}_n$.
Corollary~\ref{cor:uniqueness} is the specialization to the elliptope $\mathcal{E}_n$.

\subsection{Simplicial faces of the elliptope}
\label{sec:simplicial-faces}

As we have seen, the simplicial faces of convex bodies play a central role
in determining when convex representations are unique.  In this section,
we begin our investigation into simplicial faces of the elliptope.

\subsubsection{Schur independence}

First, recall that a set $\{ \vct{s}_1, \dots, \vct{s}_r \} \subseteq \{ \pm 1 \}^n$
of sign vectors is Schur independent if
the family $\{ \mathbf{e} \} \cup \{ \vct{s}_i \odot \vct{s}_j :1 \leq  i < j \leq r \} \subset \R^n$
is linearly independent.
It is easy to check that Schur independence implies ordinary linear independence.

\begin{lemma}[Schur independence implies linear independence] \label{lem:schur_to_linear}
A Schur-independent set of sign vectors is also linearly independent.
\end{lemma}

\begin{proof}
Let $\left\{ \vct{s}_1,\ldots,\vct{s}_r \right\} \subseteq \left\{ \pm 1 \right\}^n$
be Schur independent.  Suppose that $\lambda_1,\ldots,\lambda_r$ are real coefficients for which
$\sum_{i=1}^r \lambda_i \vct{s}_i = \vct{0}$. Since $\vct{s}_1 \odot \vct{s}_1 = \mathbf{e}$,
\begin{displaymath}
\vct{0} = \vct{s}_1 \odot \vct{0} 
	= \sum_{i=1}^r \lambda_i \vct{s}_1 \odot \vct{s}_i
	= \lambda_1 \mathbf{e} + \sum_{i=2}^r \lambda_i \vct{s}_1 \odot \vct{s}_i.
\end{displaymath}
Schur independence forces the family $\{ \mathbf{e} \} \cup \{ \vct{s}_1 \odot \vct{s}_i : 2 \leq i \leq r \}$
to be linearly independent.  We conclude that $\lambda_1=\cdots = \lambda_r = 0$.  
\end{proof}

\subsubsection{Schur independence and simplicial faces}

Laurent \& Poljak~\cite{LP96:Facial-Structure} identified the concept
of Schur independence in their work on the structure of the elliptope.
In particular, they proved that Schur independence provides a sufficient condition
for rank-one sign matrices to generate a simplicial face of the elliptope.

\begin{fact}[Laurent \& Poljak] \label{fact:LP}
Let $\mathcal{S} = \{ \vct{s}_1,\ldots,\vct{s}_r \} \subseteq \left\{ \pm 1 \right\}^n$
be a Schur-independent family of sign vectors.  Then 
$\conv \{ \vct{ss}^\transp : \vct{s} \in \mathcal{S} \}$
is a simplicial face of the elliptope $\mathcal{E}_n$.
\end{fact}

Fact~\ref{fact:LP} follows from \cite[Thm.~4.2]{LP96:Facial-Structure}
and Lemma~\ref{lem:schur_to_linear}.
Alternatively, we can establish the result using Theorem~\ref{thm:implicit},
whose proof appears below in Section~\ref{sec:separator}.

\begin{proof}[Proof of Fact~\ref{fact:LP} from Theorem~\ref{thm:implicit}]
Theorem~\ref{thm:implicit} implies that
$\mathcal{F} = \conv\{ \vct{ss}^\transp : \vct{s} \in \mathcal{S} \}$
is an exposed face of $\mathcal{E}_n$, hence it is a face.
Lemma~\ref{lem:schur_to_linear} ensures that $\mathcal{S}$
is a linearly independent set, which further implies that
$\{ \vct{ss}^{\transp} : \vct{s} \in \mathcal{S} \} \subset \Sym_n$ is
affinely independent.  Thus, $\mathcal{F}$ is a simplex.
\end{proof}

We have established the converse of Fact~\ref{fact:LP}.
In other words, the Schur independence condition is also necessary for a family
of rank-one sign matrices to generate a simplicial face of the elliptope.

\begin{lemma}[Converse of Fact~\ref{fact:LP}] \label{lem:necessary}
Let $\mathcal{S} = \{ \vct{s}_1, \dots, \vct{s}_r \} \subseteq \{ \pm 1 \}^n$ be a set of sign vectors.
If $\conv\{ \vct{ss}^\transp : \vct{s} \in \mathcal{S} \}$
is a simplicial face of the elliptope $\mathcal{E}_n$, then $\mathcal{S}$ must be Schur independent.
\end{lemma}

\begin{proof} 
Suppose that $\mathcal{F} = \conv\{ \vct{ss}^\transp : \vct{s} \in \mathcal{S} \}$
is a simplicial face of $\mathcal{E}_n$.  We argue by contradiction.

First, assume that the family $\mathcal{S}$ 
is linearly independent but not Schur independent.  Then the matrix 
$\mtx{S} = \begin{bmatrix} \vct{s}_1 & \dots & \vct{s}_r \end{bmatrix} \in \left\{ \pm 1 \right\}^{n \times r}$
has full column rank. 
Moreover, the absence of Schur independence implies that there are scalars $\theta_0$ and $\theta_{ij} = \theta_{ji}$, not all vanishing, for which
\begin{displaymath}
\theta_0 \mathbf{e} + \sum_{i \neq j} \theta_{ij} \vct{s}_i \odot \vct{s}_j = \vct{0}.
\end{displaymath}
Define a matrix $\mtx{\Theta} \in \Sym_r$ whose entries are $\left[ \mtx{\Theta} \right]_{ii}=0$ for each $i$ and $\left[ \mtx{\Theta} \right]_{ij}=\theta_{ij}$ for $i \neq j$.  For a parameter $\eps > 0$, we can introduce a pair of matrices
\begin{displaymath}
\mtx{A}_{\pm} = \mtx{S} \left( \frac{1 \pm \eps \theta_0}{r} \Id \pm \eps \mtx{\Theta} \right) \mtx{S}^\transp \in \Sym_n.
\end{displaymath}
Whenever $\eps$ is sufficiently small, both matrices $\mtx{A}_\pm$ are psd. Furthermore, by construction,
\begin{subequations}
\begin{align}
\mathrm{diag}(\mtx{A}_\pm)
&= \frac{1 \pm \eps \theta_0}{r} \sum_{i=1}^r \mathrm{diag}(\vct{s}_i \vct{s}_i^\transp) \pm \eps \sum_{i \neq j} \theta_{ij} \mathrm{diag}(\vct{s}_i \vct{s}_j^\transp ) \nonumber \\
&= \mathbf{e} \pm \eps \left( \theta_0 \mathbf{e} + \sum_{i \neq j} \theta_{ij} \vct{s}_i \odot \vct{s}_j \right) = \mathbf{e} \pm \eps \vct{0} = \mathbf{e}. \nonumber
\end{align}
\end{subequations}
In other words, both matrices $\mtx{A}_\pm$ belong to the elliptope $\mathcal{E}_n$.
Next, we verify that the average of the two matrices coincides with the barycenter of the set $\mathcal{F}$.
That is,
\begin{displaymath}
\frac{1}{2} \left(\mtx{A}_+ + \mtx{A}_- \right) = \frac{1}{r}\mtx{S}  \Id \mtx{S}^\transp = \frac{1}{r} \sum_{i=1}^r \vct{s}_i \vct{s}_i^\transp \in \mathcal{F}.
\end{displaymath}
On the other hand, neither $\mtx{A}_+$ nor $\mtx{A}_-$ is contained in $\mathcal{F}$.
To see why, just observe that the family
$\{ \vct{s}_i \vct{s}_j^\transp : 1 \leq i, j \leq r \}$
is linearly independent because $\mtx{S}$ has full column rank.  Thus,
the nonzero off-diagonal entries in $\mtx{\Theta}$
contribute to $\mtx{A}_{\pm}$ a nonzero matrix that does not belong to $\mathcal{F}$.
But this contradicts the defining property of a face, Definition~\ref{def:face}.
Indeed, $\mtx{A}_\pm \in \mathcal{E}_n$ and $\tfrac{1}{2} (\mtx{A}_+ + \mtx{A}_-) \in \mathcal{F}$,
but $\mtx{A}_\pm \notin \mathcal{F}$.

Next, assume that the family $\mathcal{S}$ of sign vectors is neither
linearly independent nor Schur independent.  Let $\mathcal{S}'$ be a maximal linearly
independent subset of $\mathcal{S}$.  Define the set
$\mathcal{F}' = \conv \{ \vct{ss}^\transp : \vct{s} \in \mathcal{S}' \}$.
Fact~\ref{fact:face-simplex} implies that $\mathcal{F}'$ is a simplicial
face of the simplex $\mathcal{F}$.  By transitivity, 
$\mathcal{F}'$ is also a simplicial face of $\mathcal{E}_n$.
On the other hand, we can repeat the argument from the last paragraph
with the simplicial face $\mathcal{F}'$ and the set $\mathcal{S}'$.
Again, we reach a contradiction.
\end{proof}

\subsubsection{Proof of Theorem~\ref{thm:simplicial_characterization}}

Theorem~\ref{thm:simplicial_characterization} summarizes the results of Fact~\ref{fact:LP}
and Lemma~\ref{lem:necessary}. There is a one-to-one correspondence between
Schur-independent families of sign vectors and simplicial
faces of the elliptope generated by rank-one sign matrices.
The second claim follows from Proposition~\ref{prop:polytope-unique}
and Fact~\ref{fact:LP}.  A matrix in the relative interior of a simplicial
face has a unique sign component decomposition.

\subsection{Explicit separators for simplicial faces of the elliptope}
\label{sec:separator}

In the last section, we developed a characterization of the simplicial faces
of the elliptope that are generated by rank-one sign matrices.  In this section,
we prove Theorem~\ref{thm:implicit}, which describes an explicit linear
functional that exposes one of these distinguished faces.
This result leads to a simple semidefinite representation for such a face.

\begin{proof}[Proof of Theorem~\ref{thm:implicit}]
Recall that $\mathcal{S} = \{ \vct{s}_1, \dots, \vct{s}_r \} \subseteq \{ \pm 1 \}^n$
is a Schur independent set of sign vectors, and $\mtx{P} \in \Sym_n$ is the orthogonal
projector onto the span of $\mathcal{S}$.
For any matrix $\mtx{X} \in \mathcal{E}_n$, 
\begin{equation} \label{eq:hoelder}
\psi(\mtx{X}) = n^{-1} \trace( \mtx{PX} )
	\leq n^{-1} \norm{ \mtx{P} }_{S_{\infty}} \norm{ \mtx{X} }_{S_1}
	= n^{-1} \trace( \mtx{X} ) = 1.
\end{equation}
We have written $\norm{ \cdot }_{S_p}$ for the Schatten $p$-norm,
and we have invoked the H{\"o}lder inequality for Schatten norms \cite[Ex.~IV.2.12]{Bha97:Matrix-Analysis}.

Next, we must verify that $\psi(\mtx{X}) = 1$ precisely when $\mtx{X}$
belongs to the set $\mathcal{F}$ described in the proposition.
Fix a matrix $\mtx{X} \in \mathcal{F}$.  It admits a decomposition as
\begin{displaymath}
\mtx{X} = \sum_{i=1}^r \tau_i \vct{s}_i \vct{s}_i^\transp
\quad\text{where}\quad
(\tau_1, \dots, \tau_r) \in \Delta_r.
\end{displaymath}
The orthogonal projector $\mtx{P}$ onto $\lspan \mathcal{S}$ clearly obeys $\mtx{P} \vct{s}_i = \vct{s}_i$
for each index $1 \leq i \leq r$.  Consequently,
\begin{displaymath}
\psi (\mtx{X}) = n^{-1} \sum_{i=1}^r \tau_i \trace \left( \mtx{P} \vct{s}_i \vct{s}_i^\transp \right)
	= n^{-1}\sum_{i=1}^r \tau_i \trace \left( \vct{s}_i \vct{s}_i^\transp \right)
	= \sum_{i=1}^r \tau_i = 1.
\end{displaymath}
This establishes $\mathcal{F} \subseteq \left\{ \mtx{X} \in \mathcal{E}_n: \psi (\mtx{X}) = 1 \right\}$.

To obtain the reverse inclusion, select a matrix $\mtx{X} \in \mathcal{E}_n$ for which $\psi(\mtx{X}) = 1$.
The relation~\eqref{eq:hoelder} holds with equality, so
$\range( \mtx{X} ) \subseteq \range( \mtx{P} ) = \lspan \mathcal{S}$.
For a symmetric matrix, the range and co-range coincide,
and we can write
\begin{displaymath}
\mtx{X} = \tfrac{1}{2} \sum_{i \leq j} \theta_{ij} \big( \vct{s}_i \vct{s}_j^\transp + \vct{s}_j \vct{s}_i^\transp \big)
\quad\text{for $\theta_{ij} \in \R$.}
\end{displaymath}
The matrix $\mtx{X}$ belongs to the elliptope, so
\begin{displaymath}
\mathbf{e} = \diag (\mtx{X})
	= \tfrac{1}{2} \sum_{i \leq j} \theta_{ij} \big( \diag \big(\vct{s}_i \vct{s}_j^\transp \big)
		+ \diag\big(\vct{s}_j \vct{s}_i^\transp \big) \big)
	= \left(\sum_{i=1}^r \theta_{ii} \right) \mathbf{e} + \sum_{i<j} \theta_{ij} \vct{s}_i \odot \vct{s}_j.
\end{displaymath}
Since $\mathcal{S}$ is Schur independent, the vectors on the right-hand side of the latter display form
a linearly independent collection.
It follows that $\sum_{i=1}^r \theta_{ii} =1$ and $\theta_{ij}=0$ whenever $i < j$.
Abbreviating $\alpha_i = \theta_{ii}$, we can express $\mtx{X} = \sum_{i=1}^r \alpha_i \vct{s}_i \vct{s}_i^\transp$
as an affine combination ($\sum_i \alpha_i = 1$) of rank-one sign matrices.

To complete the argument,
introduce the matrix $\mtx{S} = \begin{bmatrix} \vct{s}_1 &\ldots&\vct{s}_r \end{bmatrix} \in \left\{ \pm 1 \right\}^{n \times r}$,
and note that $\mtx{S}$ has full column rank because of Lemma~\ref{lem:schur_to_linear}.
We can write $\mtx{X} = \mtx{S} \, \diag(\vct{\alpha}) \, \mtx{S}^\transp$
where $\vct{\alpha} = (\alpha_1, \dots, \alpha_r)$.
Since $\mtx{X}$ belongs to the elliptope, $\mtx{X}$ is psd.
Fact~\ref{fact:psd_invariance} implies that $\diag(\vct{\alpha})$ is psd, so $\vct{\alpha}$
is nonnegative.  Therefore, $\vct{\alpha} \in \Delta_r$ is a vector of convex coefficients.  We conclude that
$\mtx{X} \in \conv \{ \vct{ss}^\transp : \vct{s} \in \mathcal{S} \} = \mathcal{F}$.
\end{proof}

\section{Computing a sign component decomposition}
\label{sec:scd-algorithm}

In the last section, we developed a geometric analysis of
the sign component decomposition~\eqref{eq:binary_factorization}
by making a connection with simplicial faces of the elliptope.
Having completed this groundwork, we can prove Theorem~\ref{thm:correct},
which states that Algorithm~\ref{alg:symfactor}
is a correct method for computing sign component decompositions.

\subsection{Step 1: Random optimization}

Our first goal is to justify the claim that random optimization
allows us to exhibit one of the rank-one sign matrix factors in the
sign component decomposition~\eqref{eq:symmetric_factorization_restatement}
of the matrix $\mtx{A}$.  We derive this conclusion from
a more general result.

\begin{lemma}[Random optimization] \label{lem:random}
Consider a family $\mathcal{U} = \{ \vct{u}_1, \ldots, \vct{u}_r \} \subset \R^n$,
in which no pair of vectors satisfies $\vct{u}_i = \pm \vct{u}_j$ when $i \neq j$.
Introduce a convex set of symmetric matrices
\begin{displaymath}
\mathcal{P} = \conv \{ \vct{uu}^\transp : \vct{u} \in \mathcal{U} \} \subset \Sym_n.
\end{displaymath}
Draw a standard normal vector $\vct{g} \in \R^n$, and construct the linear functional
$f (\vct{X}) = \vct{g}^\transp \mtx{X} \vct{g}$ for $\mtx{X} \in \Sym_n$.
Then, with probability one, there exists an index $1 \leq k \leq r$ for which
\begin{displaymath}
f(\vct{u}_k \vct{u}_k^\transp ) > f(\mtx{X}) \quad\text{for all $\mtx{X} \in \mathcal{P}$.}
\end{displaymath}
\end{lemma}

\begin{proof}
Since $\mathcal{U}$ is finite, the maximum value of $f$ over the convex hull $\mathcal{P}$
satisfies
\begin{displaymath}
\max_{\mtx{X} \in \mathcal{P}}\ f(\mtx{X})
	= \max_{\vct{\alpha} \in \Delta_r}\ \sum_{i=1}^r \alpha_i f( \vct{u}_i \vct{u}_i^\transp )
	= \max_{1 \leq i \leq r}\ f( \vct{u}_i \vct{u}_i^\transp ).
\end{displaymath}
Moreover, if $f( \vct{u}_k \vct{u}_k^\transp ) > f( \vct{u}_i \vct{u}_i^\transp)$
for all $i \neq k$, then the maximum on the left-hand side is attained uniquely
at the matrix $\mtx{X} = \vct{u}_k \vct{u}_k^\transp$.

It suffices to prove that, with probability one,
the linear functional $f$ takes distinct values at the rank-one matrices
$\vct{uu}^\transp$ given by $\vct{u} \in \mathcal{U}$.
First, observe that
$f( \vct{uu}^\transp ) = \ip{ \vct{g} }{ \vct{u} }^2$.
A short calculation reveals that
\begin{displaymath}
f( \vct{uu}^\transp ) = f( \vct{vv}^\transp )
\quad\text{if and only if}\quad
\ip{ \vct{g} }{ \vct{u} + \vct{v} } = 0
\quad\text{or}\quad \ip{ \vct{g} }{ \vct{u} - \vct{v} } = 0.
\end{displaymath}
By rotational invariance, 
each of the inner products follows a normal distribution:
\begin{displaymath}
\ip{ \vct{g} }{ \vct{u} + \vct{v} }
	\sim \textsc{normal}\big( 0, \norm{ \vct{u} + \vct{v} }_{\ell_2}^2 \big)
	\quad\text{and}\quad
\ip{ \vct{g} }{ \vct{u} - \vct{v} }
	\sim \textsc{normal}\big( 0, \norm{ \vct{u} - \vct{v} }_{\ell_2}^2 \big).	
\end{displaymath}
Unless $\vct{v} = \pm \vct{u}$, neither variance can vanish.
As a consequence, we may evaluate the probability
\begin{displaymath}
\Prob{ \text{$\ip{ \vct{g} }{ \vct{u}_i }^2 = \ip{ \vct{g} }{ \vct{u}_j }^2$ for some $i \neq j$} }
	\leq \sum_{i < j} \left( \Prob{ \ip{ \vct{g} }{ \vct{u}_i + \vct{u}_j } = 0 } 
	+ \Prob{ \ip{ \vct{g} }{ \vct{u}_i - \vct{u}_j } = 0 } \right)
	= 0.
\end{displaymath}
The last relation holds because $\vct{u}_i$ never coincides with $\pm \vct{u}_j$ for $i < j$.
Take the complement of this event to reach the conclusion.
\end{proof}

\subsection{Step 2: Deflation}

Random optimization allows us to identify a single sign component
in the decomposition~\eqref{eq:symmetric_factorization_restatement}.
In order to iterate, we must remove the contribution of this sign
component from the matrix that we are factoring.
The following general result shows how to extract a rank-one factor from a psd matrix,
leaving a convex combination of the other rank-one factors.

\begin{lemma}[Deflation] \label{lem:deflation}
Consider a linearly independent family $\mathcal{U} = \{ \vct{u}_1, \ldots, \vct{u}_r \} \subset \R^n$,
and suppose that
\begin{displaymath}
\mtx{M} = \sum_{i=1}^r \alpha_i \vct{u}_i \vct{u}_i^\transp
\quad\text{where}\quad
\vct{u}_i \in \mathcal{U}
\quad\text{and}\quad
(\alpha_1, \dots, \alpha_r) \in \Delta_r^+.
\end{displaymath}
Fix an index $1 \leq k \leq r$, and consider the semidefinite program
\begin{displaymath}
\underset{\zeta \in \R}{\maximize}\quad \zeta
\quad\subjto\quad
\zeta \mtx{M} + (1 - \zeta) \vct{u}_k \vct{u}_k^\transp \psdge \mtx{0}.
\end{displaymath}
For the unique solution $\zeta_{\star} = (1 - \alpha_k)^{-1}$, it holds that
\begin{displaymath}
\mtx{M}' = \zeta_{\star} \mtx{M} + (1 - \zeta_{\star}) \vct{u}_k \vct{u}_k^\transp
	= \sum_{i \neq k} \frac{\alpha_i}{1 - \alpha_k} \vct{u}_i \vct{u}_i^\transp
	\in \relint \conv \big\{ \vct{u}_i \vct{u}_i^\transp : \text{$1 \leq i \leq r$ and $i \neq k$} \big\}.
\end{displaymath}
\end{lemma}

\begin{proof}
Without loss of generality, assume that $k = 1$.  Since $\mathcal{U}$ is linearly independent,
the matrix $\mtx{U} = \begin{bmatrix} \vct{u}_1 & \dots & \vct{u}_r \end{bmatrix}$ has full
column rank.  In turn, the conjugation rule (Fact~\ref{fact:psd_invariance}) implies that
\begin{displaymath}
\zeta \mtx{M} + (1 - \zeta) \vct{u}_1 \vct{u}_1^\transp
	= \mtx{U} \, \diag \big( \alpha_1 \zeta + (1 - \zeta), \alpha_2 \zeta, \dots, \alpha_r \zeta \big) \, \mtx{U}^\transp \psdge \mtx{0}
\end{displaymath}
if and only if the diagonal matrix is psd.  Equivalently, $\zeta$ is feasible if and only if
$0 \leq \zeta \leq (1- \alpha_1)^{-1}$.
The optimal point $\zeta_{\star}$ for the semidefinite program saturates the upper bound.
The second claim follows readily from a direct computation.
\end{proof}

\subsection{Proof of Theorem~\ref{thm:correct}}
\label{sec:step-iii}

We are now prepared to prove Theorem~\ref{thm:correct},
which states that Algorithm~\ref{alg:symfactor} is correct.
The argument is based on induction on the rank of the input matrix.

First, suppose that $\mtx{A} = \vct{ss}^\transp$ is a rank-one correlation
matrix generated by a sign vector $\vct{s} \in \{ \pm 1 \}^n$.
In this case, the sign component decomposition of $\mtx{A}$ is already manifest.
By factorizing $\mtx{A}$, we obtain the computed sign component $\pm \vct{s}$.

Now, for $r \geq 2$, suppose that $\mtx{A}$ is a rank-$r$ correlation matrix
with sign component decomposition
\begin{equation} \label{eqn:scd-alg-proof}
\mtx{A} = \sum_{i=1}^r \tau_i \vct{s}_i \vct{s}_i^\transp
\quad\text{for $\vct{s}_i \in \{ \pm 1 \}^n$ and $(\tau_1, \dots, \tau_r) \in \Delta_r^+$.}
\end{equation}
We assume that $\mathcal{S} = \{ \vct{s}_1, \dots, \vct{s}_r \}$ is Schur independent.
Lemma~\ref{lem:schur_to_linear} states that $\mathcal{S}$ is linearly independent.
In particular, $\vct{s}_i \neq \pm \vct{s}_j$ when $i \neq j$.  Moreover,
Fact~\ref{fact:LP} ensures that $\mathcal{F} = \{ \vct{ss}^\transp : \vct{s} \in \mathcal{S} \}$
is a simplicial face of the elliptope that contains the matrix $\mtx{A}$
in its relative interior.

Compute the orthogonal projector $\mtx{P} \in \Sym_n$ onto $\range(\mtx{A}) = \lspan \mathcal{S}$,
and define the linear functional $\psi(\mtx{X}) = n^{-1} \trace(\mtx{PX})$
that exposes the face $\mathcal{F}$.
Draw a standard normal vector $\vct{g} \in \R^n$.  Find the solution $\mtx{X}_{\star}$
to the semidefinite program
\begin{equation} \label{eqn:random-opt-pf}
\underset{\mtx{X} \in \Sym_n}{\maximize} \quad \vct{g}^\transp \mtx{X} \vct{g}
	\quad\subjto\quad \text{$\psi(\mtx{X}) = 1$ and  $\mtx{X} \in \mathcal{E}_n$.}
\end{equation}
According to Theorem~\ref{thm:implicit},
the feasible set of this optimization problem is precisely the simplicial face $\mathcal{F}$
that contains $\mtx{A}$.
An application of Lemma~\ref{lem:random} shows that the optimal point is unique
with probability one, and $\mtx{X}_{\star} = \vct{s}_k \vct{s}_k^\transp$ for
some index $1 \leq k \leq r$.  By factorizing $\mtx{X}_{\star}$, we compute one
sign component $\pm \vct{s}_k$.  This justifies Step 1 of Algorithm~\ref{alg:symfactor}.

Next, we find the unique solution $\zeta_{\star}$ to the semidefinite program
\begin{equation} \label{eqn:deflation-pf}
\underset{\zeta \in \R}{\maximize}\quad \zeta
\quad\subjto\quad
\zeta \mtx{A} + (1 - \zeta) \mtx{X}_{\star} \psdge \mtx{0}.
\end{equation}
Lemma~\ref{lem:deflation} shows that $\zeta_{\star} = (1 - \tau_k)^{-1}$,
where $\tau_k$ is the coefficient associated with $\vct{s}_k \vct{s}_k^\transp$
in the representation~\eqref{eqn:scd-alg-proof} of the matrix $\mtx{A}$.
Moreover, we can form the matrix
\begin{equation} \label{eqn:scd-A'}
\mtx{A}' = \zeta_{\star} \mtx{A} + (1 - \zeta_{\star}) \mtx{X}_{\star}
	= \sum_{i \neq k} \frac{\tau_i}{1 - \tau_k} \vct{s}_i \vct{s}_i^\transp
	=: \sum_{i \neq k} \tau_i' \vct{s}_i \vct{s}_i^\transp
	\quad\text{where}\quad
	\vct{\tau}' \in \Delta_{r-1}^+.
\end{equation}
Recall that every subset of a Schur-independent set remains Schur independent.
Therefore, Step 2 of Algorithm~\ref{alg:symfactor} produces a correlation matrix $\mtx{A}'$
with rank $r-1$ that admits a sign component decomposition~\eqref{eqn:scd-A'}
whose sign components form a Schur-independent family.

By induction, we can apply the same procedure to compute the sign
components of the matrix $\mtx{A}'$ defined in~\eqref{eqn:scd-A'}.
This justifies the iteration procedure, Step 3 in Algorithm~\ref{alg:symfactor}.

Now, suppose that $\{\tilde{\vct{s}}_1, \dots, \tilde{\vct{s}}_r\} \subseteq \{\pm 1\}^n$
is the set of sign components computed by this iteration.  There is a permutation $\pi$
such that $\vct{s}_{\pi(i)} = \xi_i \tilde{\vct{s}}_i$
and $\xi_i \in \{ \pm 1 \}$ for each index $i = 1, \dots, r$.
To determine the convex coefficients in the sign component decomposition of $\mtx{A}$,
we find the solution $\tilde{\vct{\tau}} \in \R^r$ to the linear system
\begin{displaymath}
\mtx{A} = \sum_{i=1}^r \tilde{\tau}_i \tilde{\vct{s}}_i \tilde{\vct{s}}_i^\transp.
\end{displaymath}
The computed sign components must be linearly independent (since the original sign components
are linearly independent), so the linear system has a unique solution.
In view of~\eqref{eqn:scd-alg-proof}, it must be the case that $\tau_{\pi(i)} = \tilde{\tau}_i$
for each index $i$.  In other words, $\{ (\tilde{\tau}_i, \tilde{\vct{s}}_i ) : i = 1, \dots, r \}$
is a parametric representation of the sign component decomposition of $\mtx{A}$.
This justifies Step 4 of Algorithm~\ref{alg:symfactor},
and the proof is complete.

\begin{remark}[Accuracy]
Since the sign components are discrete, we can identify each one by
solving the random optimization problem~\eqref{eqn:random-opt-pf}
with rather limited accuracy.  In contrast, to remove the sign
component completely, we should solve the deflation
problem~\eqref{eqn:deflation-pf} to high accuracy.
The deflation step~\eqref{eqn:deflation-pf}
can be rewritten as a generalized eigenvalue problem,
which makes this task routine.
\end{remark}

\begin{remark}[Dimension reduction]
As it is stated, Algorithm~\ref{alg:symfactor} requires us to solve semidefinite programs
in an $n \times n$ matrix variable.  It is not hard to develop an equivalent
procedure based on optimization over a much lower-dimensional space of matrices.
This approach has significantly lower resource usage.
For the sake of brevity, we omit its discussion here and refer to the
appendix for details.
\end{remark}

\section{Binary component decomposition} \label{sec:01factorization}

In this section, we develop a procedure (Algorithm~\ref{alg:binfactor})
for binary component decomposition, and we prove that it succeeds under
a Schur independence condition (Theorem~\ref{thm:bcd-main}).
Our approach reduces the problem of computing a binary component
decomposition to the problem of computing a sign component decomposition.

\subsection{Correspondence between binary vectors and sign vectors}

Recall that we can place sign vectors and binary vectors in one-to-one
correspondence via the affine map
\begin{displaymath}
\mtx{F} : \{ 0, 1 \}^n \to \{ \pm 1 \}^n  
\quad\text{where}\quad
\mtx{F} : \vct{z} \mapsto 2\vct{z} - \mathbf{e}
\quad\text{and}\quad
\mtx{F}^{-1} : \vct{s} \mapsto \tfrac{1}{2} (\vct{s} + \mathbf{e}).
\end{displaymath}
The correspondence between sign component decompositions and binary component decompositions,
however, is more subtle because they are invariant under different symmetries.  Indeed,
$\vct{ss}^\transp$ is invariant under flipping the sign of $\vct{s} \in \{ \pm 1 \}^n$,
while $\vct{zz}^\transp$ is uniquely determined for each $\vct{z} \in \{0, 1\}^n$.

\subsection{Reducing binary component decomposition to sign component decomposition}

Given a matrix that has a binary component decomposition, we can solve a linear system
to obtain a matrix that has a closely related sign component decomposition.

\begin{proposition}[Binary component decomposition: Reduction] \label{prop:bcd-scd}
Consider a matrix $\mtx{H} \in \Sym_n$ that has a binary component decomposition
\begin{equation} \label{eqn:bcd-reduction}
\mtx{H} = \sum_{i=1}^r \tau_i \vct{z}_i \vct{z}_i^\transp
\quad\text{for $\vct{z}_i \in \{0, 1\}^n$ and $(\tau_1,\dots,\tau_r) \in \Delta_r^+$.}
\end{equation}
Define the correlation matrix $\mtx{A} \in \mathcal{E}_n$ with sign component decomposition
\begin{displaymath}
\mtx{A} = \sum_{i=1}^r \tau_i \vct{s}_i \vct{s}_i^\transp
\quad\text{where $\vct{s}_i = \mtx{F}(\vct{z}_i)$ for each $i$.}
\end{displaymath}
Then $\mtx{A}$ is the unique solution to the linear system
\begin{equation} \label{eqn:bcd-to-scd}
\diag(\mtx{X}) = \mathbf{e}
\quad\text{and}\quad
\mtx{R}(  4\mtx{H} - \mtx{X} ) \mtx{R} = \mtx{0}
\quad\text{where $\mtx{X} \in \Sym_n$.}
\end{equation}
Here, $\mtx{R} = \Id - n^{-1} \mathbf{e}\mathbf{e}^\transp$ denotes the orthogonal projector
onto $\{ \mathbf{e} \}^\perp \subset\R^n$.
\end{proposition}

\begin{proof}
For a binary vector $\vct{z} \in \{ 0, 1 \}^n$, the sign vector $\vct{s} = \mtx{F}(\vct{z})$
satisfies the identity
$\vct{ss}^\transp = (2\vct{z} - \mathbf{e})(2\vct{z} - \mathbf{e})^\transp$.
The projector $\mtx{R}$ annihilates the vector $\mathbf{e}$, so we can conjugate by $\mtx{R}$ to obtain
$\mtx{R}\vct{s}\vct{s}^\transp \mtx{R} = 4 \mtx{R}\vct{z}\vct{z}^\transp \mtx{R}$.
Instantiate this relation for each of the vectors $\vct{z}_i$ that appears
in the binary component decomposition~\eqref{eqn:bcd-reduction},
and average using the weights $(\tau_1,\dots,\tau_r) \in \Delta_r^+$ to arrive at
\begin{displaymath}
\mtx{RAR}
	= \sum_{i=1}^n \tau_i \mtx{R}\vct{s}_i\vct{s}_i^\transp \mtx{R}
	= 4 \sum_{i=1}^n \tau_i \mtx{R}\vct{z}_i\vct{z}_i^\transp \mtx{R}
	= 4 \mtx{RHR}.
\end{displaymath}
The correlation matrix $\mtx{A}$ has a unit diagonal,
so it solves the linear system~\eqref{eqn:bcd-to-scd}.

We need to confirm that $\mtx{A}$ is the only solution to~\eqref{eqn:bcd-to-scd}.
The kernel of the linear map $\mtx{X} \mapsto \mtx{RXR}$ on $\Sym_n$ consists of
matrices with the form $\mathbf{e}\vct{x}^\transp + \vct{x}\mathbf{e}^\transp$ for $\vct{x} \in \R^n$.
Therefore, we can parameterize each solution $\mtx{X}$ of the second equation in~\eqref{eqn:bcd-to-scd}
as $\mtx{X} = \mtx{A} + \mathbf{e}\vct{x}^\transp + \vct{x}\mathbf{e}^\transp$.
But the first equation in~\eqref{eqn:bcd-to-scd} requires that
\begin{displaymath}
\mathbf{e} = \diag(\mtx{X}) = \diag(\mtx{A}) + \diag(\mathbf{e} \vct{x}^\transp) + \diag(\vct{x}\mathbf{e}^\transp)
	= \mathbf{e} + 2 \vct{x}.
\end{displaymath}
Therefore, $\vct{x} = \vct{0}$, and so $\mtx{A}$ is the only matrix that
solves both equations.
\end{proof}

\subsection{Resolving the sign ambiguity}

Proposition~\ref{prop:bcd-scd} shows that we can replace the matrix $\mtx{H}$
by a correlation matrix $\mtx{A}$ whose sign components are related to
the binary components in $\mtx{H}$.  Let us explain how to resolve
the sign ambiguity in the sign components of $\mtx{A}$ to identify the
correct binary components for $\mtx{H}$.

\begin{proposition}[Sign ambiguity] \label{prop:bcd-sign}
Instate the notation of Proposition~\ref{prop:bcd-scd}.
Assume that the correlation matrix $\mtx{A}$ has a unique
sign component decomposition with parametric representation
$\{ (\tau_i, \tilde{\vct{s}}_i) : i = 1, \dots, r \}$,
and assume that the sign components form a linearly independent family.
Find the unique solution $\vct{\xi} \in \R^r$ to the linear system
\begin{displaymath}
n \sum_{i=1}^r  \tau_i \xi_i \tilde{\vct{s}}_i
	= (4 \mtx{H} - \mtx{A}) \mathbf{e} - 2n \trace(\mtx{H}) \mathbf{e}.
\end{displaymath}
Then the binary components of $\mtx{H}$ are given by
$\vct{z}_i = \frac{1}{2}(\xi_i \tilde{\vct{s}}_i + \mathbf{e})$ for each $1 \leq i \leq r$.
\end{proposition}

\begin{proof}
For a binary vector $\vct{z} \in \{0,1\}^n$, define $\vct{s} = \mtx{F}(\vct{z}) = 2 \vct{z} - \mathbf{e}$.
By direct computation, 
\begin{displaymath}
\vct{s} \mathbf{e}^\transp = 4 \vct{zz}^\transp - 2 \mathbf{e} \vct{z}^\transp - \vct{ss}^\transp.
\end{displaymath}
Right-multiply the last display by the vector $\mathbf{e}$ to arrive at
\begin{displaymath}
n\vct{s}
	= 4 \vct{zz}^\transp \mathbf{e} - 2n (\vct{z}^\transp \mathbf{e}) \mathbf{e} - \vct{ss}^\transp \mathbf{e} 
	= 4 \vct{zz}^\transp \mathbf{e} - 2n \trace(\vct{zz}^\transp) \mathbf{e} - \vct{ss}^\transp \mathbf{e}.
\end{displaymath}
The last relation holds because a 0--1 vector $\vct{z}$ satisfies $\vct{z}^\transp \mathbf{e} = \trace(\vct{zz}^\transp)$.
Instantiate the last display for the vectors $\vct{s}_i = \mtx{F}(\vct{z}_i)$,
and form the average using the weights $(\tau_1,\dots,\tau_r) \in \Delta_r^+$ to obtain
\begin{displaymath}
n \sum_{i=1}^r \tau_i \vct{s}_i
	= 4 \mtx{H} \mathbf{e} - 2n \trace(\mtx{H}) \mathbf{e} - \mtx{A} \mathbf{e}.
\end{displaymath}
We have used the definitions of $\mtx{H}$ and $\mtx{A}$ from the statement of Proposition~\ref{prop:bcd-scd}.

By uniqueness, the sign components $\tilde{\vct{s}}_i$ in the parametric representation
coincide with the vectors $\vct{s}_i$ up to global sign flips;
that is, $\vct{s}_i = \xi_i \tilde{\vct{s}}_i$ where $\xi_i \in \{ \pm 1 \}$ for each index $i$.
Substitute this relation into the last display to obtain
\begin{displaymath}
n \sum_{i=1}^r \tau_i \xi_i \tilde{\vct{s}}_i
	= 4 \mtx{H} \mathbf{e} - 2n \trace(\mtx{H}) \mathbf{e} - \mtx{A} \mathbf{e}.
\end{displaymath}
This is a consistent linear system in the variables $\xi_1, \dots, \xi_r$.
The solution is unique because $\{ \tilde{\vct{s}}_1, \dots, \tilde{\vct{s}}_r \}$
is a linearly independent family.
Therefore, we can obtain the sign pattern by solving the linear system,
and
\begin{displaymath}
\vct{z}_i = \mtx{F}^{-1}(\vct{s}_i) =
	\mtx{F}^{-1}(\xi_i \tilde{\vct{s}}_i) = \tfrac{1}{2} (\xi_i \tilde{\vct{s}}_i + \mathbf{e}).
\end{displaymath}
This observation completes the argument. 
\end{proof}

\subsection{Computation of the binary component decomposition}
\label{sec:bcd-compute}

Proposition~\ref{prop:bcd-scd} and Proposition~\ref{prop:bcd-sign} give us a mechanism
for computing a binary component decomposition, provided that an associated matrix has
a unique sign component decomposition with linearly independent sign components.
We can exploit our theory on the tractable computation of sign component decompositions
to identify situations where we can compute binary component decompositions.

\begin{proposition}[Schur independence: Equivalence] \label{prop:schur-bcd-scd}
A family $\{ \vct{z}_1, \dots, \vct{z}_r \} \subseteq \{0,1\}^n$
of binary vectors is Schur independent if and only if the associated family
$\{ \mathbf{e}, \mtx{F}(\vct{z}_1), \dots, \mtx{F}(\vct{z}_r) \} \subseteq \{\pm 1\}^n$
of sign vectors is Schur independent.
\end{proposition}

\begin{proof}
Set $\vct{z}_0 = \mathbf{e}$ and $\vct{s}_0 = \mathbf{e}$.  For $1 \leq i \leq r$,
define $\vct{s}_i = \mtx{F}(\vct{z}_i) = 2 \vct{z}_i - \mathbf{e}$.  Then
\begin{displaymath}
\lspan\{ \vct{z}_i \odot \vct{z}_j : 0 \leq i, j \leq r \}
	= \lspan\{ \vct{s}_i \odot \vct{s}_j : 0 \leq i, j \leq r \}.
\end{displaymath}
This point follows easily from the definition of the linear hull.
\end{proof}

With this result at hand, we can prove Theorem~\ref{thm:bcd-main}.

\begin{proof}[Proof of Theorem~\ref{thm:bcd-main}]
Suppose that $\mtx{H} \in \Sym_n$ has a binary component decomposition
$\mtx{H} = \sum_{i=1}^r \tau_i \vct{z}_i \vct{z}_i^\transp$
involving a Schur independent family $\{ \vct{z}_1, \dots, \vct{z}_r \}$
of binary components.  Introduce the associated sign vectors $\vct{s}_i = \mtx{F}(\vct{z}_i)$.
By Proposition~\ref{prop:schur-bcd-scd}, the family $\{ \vct{s}_1, \dots, \vct{s}_r \}$
of sign vectors is Schur independent, hence linearly independent by Lemma~\ref{lem:schur_to_linear}.

Proposition~\ref{prop:bcd-scd} shows that we can form the correlation matrix
$\mtx{A} = \sum_{i=1}^r \tau_i \vct{s}_i \vct{s}_i^\transp$ by solving
a linear system.  By Theorem~\ref{thm:correct}, Algorithm~\ref{alg:symfactor}
allows us to compute pairs $(\tilde{\tau}_i, \tilde{\vct{s}}_i)$ with the property
that $\tau_{\pi(i)} = \tilde{\tau}_i$ and $\vct{s}_{\pi(i)} = \xi_i \tilde{\vct{s}}_i$
where $\pi$ is a permutation and $\xi_i \in \{\pm1\}$ for each $i$.
Proposition~\ref{prop:bcd-sign} shows that we can use the computed sign
components to find the associated binary components $\vct{z}_{\pi(i)}$
that participate in $\mtx{H}$.
\end{proof}

\section{Application: activity detection in massive MIMO systems}
\label{sec:mimo}

In this section, we outline a stylized application of binary
component decomposition in modern communications.

\subsection{Motivation}

Massive connectivity is predicted to be a key feature in future wireless cellular networks (IoT) and Device-to-Device communication (D2D). Base stations will face the challenge of connecting a large number of devices and distributing communication resources  accordingly. While this seems daunting in general, a key feature of these systems is parsimony. Individual device activity is typically sporadic. 
This feature can be exploited by a two-phase approach:
\begin{enumerate}
\item \emph{Activity detection:} identify the (small) set of active users at a given time.
\item \emph{Scheduling:} distribute communication resources among these active users. 
\end{enumerate}
Recent works have pointed out that multiple antennas at the base station may help to tackle the first phase \cite{LY18:MIMO, CSY18:MIMO-AMP, HJC18:Activity-Detection}. 
The mathematical motivation behind this approach is covariance estimation. Massive MIMO systems allow for estimating the covariance matrix of an incoming signal, rather than the signal itself. 
As detailed below, this reduces the task of identifying active devices to a matrix factorization problem. The covariance matrix is proportional to a convex combination of structured rank-one factors. Each of these factors is in one-to-one correspondence with a single active device. 
Identifying this factorization in turn allows for solving the activity detection problem.

\subsection{Signal model}

Suppose that a network contains $N$ different devices and a single base station. The base station contains $M$ different antennas. Each of these antennas is capable of resolving $n$-dimensional signals. 
To perform activity detection, unique pilot sequences are distributed among the devices. Denote them by $\vct{a}_1,\ldots,\vct{a}_N \in \C^n$.
If device $k$ wants to indicate activity, it transmits its pilot $\vct{a}_k$ over the shared network.
At a given time, the base station receives noisy super-positions of several pilot sequences that passed through wireless channels. 
The channel connecting device $k$ ($1 \leq k \leq N$) with the $i$-th antenna ($1 \leq i \leq M$)
is modeled by a large-scale fading coefficient $\tau_k >0$ that is constant over all antennas and a channel vector $\bar{\vct{h}}_k \in \C^M$ that subsumes fluctuations between antennas:
\begin{displaymath}
\vct{y}_i = \sum_{k \in \mathcal{A}} \sqrt{\tau_k} \left[ \bar{\vct{h}}_k \right]_i \vct{a}_k + \vct{\eps}_i \in \C^n \quad \textrm{for}\quad  1 \leq i \leq M.
\end{displaymath}
The set $\mathcal{A} \subset \left\{1,\ldots,N \right\}$ denotes the sub-set of active devices and $\vct{\eps}_i \in \mathbb{C}^m$ represents additive noise corruption affecting the $i$-th antenna.
Simplifying assumptions, such as \emph{white noise corruption} (each $\vct{\eps}_i$ is a complex standard Gaussian vector with variance $\eps$) and \emph{spatially white channel vectors} (each $\vct{h}_k$ contains i.i.d.\ standard normal entries) imply the following simple formula for the covariance:
\begin{displaymath}
\mathrm{Cov} (\vct{y}_i) = \mathbb{E} \left[ \vct{y} \vct{y}^* \right] = \sum_{k \in \mathcal{A}} \tau_k \vct{a}_k \vct{a}_k^* + \eps \Id \quad \textrm{for all} \quad 1 \leq i \leq M.
\end{displaymath}
The MIMO setup allows for empirically approximating this covariance:
\begin{equation} \label{eq:covariance_matrix}
\mtx{Y} = M^{-1} \sum_{i=1}^M \vct{y}_i \vct{y}_i^* \overset{M \to \infty}{\longrightarrow} \sum_{k \in \mathcal{A}} \tau_k \vct{a}_k \vct{a}_k^*+\eps \Id.
\end{equation}
The (quick) rate of convergence can be controlled using matrix-valued concentration inequalities \cite{Tro12:Tail-Bounds}. 
We refer to \cite{HJC18:Activity-Detection} for a more detailed analysis and justification of the simplifying assumptions.

\subsection{Compressed activity detection via sign component decomposition}
Let $N$ be the total number of devices in the network. Set the internal dimension to
$
n = \lceil \log_2 (N) \rceil~+~1
$.
Equip each device with a unique pilot sequence $\vct{a}_i=\vct{s}_i \in \left\{ \pm 1 \right\}^n$ such that $\vct{s}_i \neq \pm \vct{s}_j$ for all $i \neq j$. 
Next, assume that the base station contains sufficiently many antennas to  accurately estimate the signal covariance matrix \eqref{eq:covariance_matrix} at any given time:
\begin{displaymath}
\mtx{Y} = \sum_{k \in \mathcal{A}} \tau_k \vct{s}_k \vct{s}_k^T + \eps \Id.
\end{displaymath} 
Standard techniques allow for removing the isotropic noise distortion $\eps \Id$. The remainder is proportional to a correlation matrix: $\bar{\mtx{Y}} = \sum_{k \in \mathcal{A}} \tau_k \vct{s}_k \vct{s}_k^T$. 
The activity pattern $\mathcal{A}$ is encoded in the sign components (pilots) of this correlation matrix. Apply Algorithm~\ref{alg:symfactor} to identify them.

Theorem~\ref{thm:scd-main} asserts that this identification succeeds, provided that the participating sign components are Schur-independent. 
This assumption imposes stringent constraints on the maximum number of active devices that can be resolved correctly, see Eq.~\eqref{eq:factorization_rank}. But beneath this threshold, Schur independence is generic. Fact~\ref{fact:schur-independence-generic} asserts that almost all activity patterns produce Schur independent pilot sequences.

The method proposed here is conceptually different from existing approaches. These assign random pilot sequences and exploit sparsity in the activity pattern -- viewed as a binary vector in $\mathbb{R}^N$ -- either via approximate message passing \cite{CSY18:MIMO-AMP} or ideas from compressed sensing \cite{HJC18:Activity-Detection,FJ19:Khatri-Rao-RIP}. 
In contrast, activity detection via sign component decomposition assigns deterministic pilot sequences that are guaranteed to work for most parsimonious activity patterns. The algorithmic reconstruction cost scales polynomially in $n \simeq \log (N)$, an exponential improvement over existing rigorous reconstruction techniques \cite{FJ19:Khatri-Rao-RIP}.

The arguments presented here are based on several idealizations and should be viewed as a proof of concept. We intend to address concrete implementations of MIMO activity detection via sign component decomposition in future work.

\section{Related work}
\label{sec:related-work}

The goal of matrix factorization is to produce a representation
of a matrix $\mtx{B} \in \R^{n \times m}$ as a product of structured
matrices.  The simplest formulation expresses 
\begin{equation} \label{eqn:two-factors}
\mtx{B} = \mtx{VW}^\adj + \mtx{E}
\quad\text{where}\quad
\mtx{V} \in \R^{n \times r}
\quad\text{and}\quad
\mtx{W} \in \R^{r \times m}.
\end{equation} 
The matrix $\mtx{E} \in \R^{n \times m}$ collects the approximation error;
the factorization is \emph{exact} if $\mtx{E} = \mtx{0}$.
It is also common to normalize the factors and expose the scaling
by means of a separate diagonal factor:
\begin{equation} \label{eqn:three-factors}
\mtx{B} = \mtx{V} \, \diag(\vct{\lambda}) \, \mtx{W}^\adj + \mtx{E}
\quad\text{where}\quad
\vct{\lambda} \in \R_+^r.
\end{equation}
We can try to expose different types of structure in the matrix
by placing appropriate constraints on the factors $\mtx{V}$
and $\mtx{W}$.  The shape of the factors may vary,
depending on the application.

The basic questions about matrix factorization are
existence, uniqueness, stability, and computational tractability.
Surprisingly, there is little rigorous theory
about matrix factorizations beyond the most classical examples.
Furthermore, a majority of the algorithmic work
consists of heuristic nonconvex optimization procedures.
The aim of this section is to summarize the literature on
discrete factorizations, as well as some general
computational approaches to matrix factorization.

\subsection{Integer factorizations}

In 1851, Hermite developed an integer analog of the
reduced row echelon form~\cite{Her51:Normal-Form}.
For an integer matrix $\mtx{B}$, the Hermite normal
form is an exact factorization~\eqref{eqn:two-factors}
where $\mtx{V}$ is a square unimodular%
\footnote{A unimodular matrix has determinant one.}
integer matrix and $\mtx{W}$ is a triangular integer matrix.
This decomposition is a discrete analog of the \textsf{QR} factorization.

Similar in spirit, the Smith normal form~\cite{Smi61:Smith-Normal}
of an integer matrix $\mtx{B}$ is an exact factorization~\eqref{eqn:three-factors}
where $\mtx{V}$ and $\mtx{W}$ are square unimodular integer matrices,
and $\vct{\lambda} \in \mathbb{N}^r$ is an integer vector with
the divisibility property $\lambda_{i+1} \, \vert \, \lambda_i$ for each $i$.
This is a discrete analog of the SVD.

Both these decompositions can be extended to a matrix whose entries
are drawn from a principal ideal domain.  For example,
with respect to the finite field $\Z_2 = \{0, 1\}$,
these normal forms lead to binary factorizations
of a binary matrix.

Both the Hermite and Smith normal forms of an integer matrix
can be computed in strongly polynomial time~\cite{KB79:Polynomial-Algorithms}.
Contemporary applications include multidimensional signal processing,
lattice computations, and solving Diophantine equations~\cite{Yap00:Fundamental-Problems}.

\subsection{Semidiscrete factorizations}

Kolda~\cite{Kol98:Phd-Thesis} coined the term \emph{semidiscrete factorization}
to describe the class of factorizations of the form~\eqref{eqn:three-factors}
where the outer factors $\mtx{V}, \mtx{W}$ are discrete
while the diagonal vector $\vct{\lambda}$ takes real values.
The literature contains several instances.

\subsubsection{Integer factorizations}

Tropp~\cite{Tro15:Integer-Factorization} proved that every positive-definite
matrix $\mtx{B}$ admits an exact semidiscrete factorization~\eqref{eqn:three-factors}
where $\mtx{V} = \mtx{W}$ and the factor $\vct{V}$ has integer entries that are
bounded in terms of the condition number and the dimension.
This result has applications in probability theory~\cite{BLX18:Multivariate-Approximation},
but the proof is nonconstructive.

\subsubsection{Ternary factorizations}

Motivated by applications in image processing,
O'Leary and Peleg~\cite{LP83:Digital-Compression}
considered an approximate semidiscrete factorization~\eqref{eqn:three-factors}
where $\mtx{V} \in \{0, \pm 1\}^{n \times r}$ and $\mtx{W} \in \{0, \pm 1\}^{m \times r}$
take ternary values.
They proposed a heuristic method for computing
the semidiscrete factorization.  At each step, they aim to
solve the integer optimization problem
\begin{equation} \label{eqn:ternary-opt}
\underset{\vct{x} \in \{0, \pm 1\}^n, \vct{y} \in \{0, \pm 1\}^m}{\maximize} \quad
	\vct{x}^\adj \mtx{B} \vct{y}.
\end{equation}
Given an approximate solution $(\vct{x}, \vct{y})$ to~\eqref{eqn:ternary-opt},
they update the target matrix as
\begin{equation} \label{eqn:deflation-classic}
\mtx{B} \mapsto \mtx{B} - \lambda \, \vct{xy}^\adj
\quad\text{where}\quad
\lambda = \frac{\vct{x}^\adj \mtx{B} \vct{y}}{\norm{\vct{x}} \norm{\smash{\vct{y}}}}.
\end{equation}
This process is sometimes known as \emph{deflation}.
It leads directly to a factorization of the form~\eqref{eqn:three-factors}.

Since it is computationally hard to solve
the optimization problem~\eqref{eqn:ternary-opt} exactly,
O'Leary and Peleg resort to alternating minimization.
They fix $\vct{x}$ and minimize with respect to $\vct{y}$;
they fix $\vct{y}$ and minimize with respect to $\vct{x}$;
and repeat.  Kolda~\cite[Prop.~6.2]{Kol98:Phd-Thesis}
later showed that these heuristics produce a sequence
of approximations with decreasing errors; there is no
control on the rate of convergence.

\subsubsection{Binary factorizations, or cut decompositions}

The \emph{cut norm} of a matrix $\mtx{B}$ is the value
of the optimization problem
\begin{equation} \label{eqn:binary-opt}
\underset{\vct{x} \in \{0,1\}^n, \vct{y} \in \{0, 1\}^m}{\maximize} \quad
	\vct{x}^\adj \mtx{B} \vct{y}.
\end{equation}
The cut norm has applications in graph theory and theoretical algorithms.

In the late 1990s, Frieze and Kannan~\cite{FK99:Cut-Decomposition} proposed
the \emph{cut decomposition}, an approximate
semidiscrete factorization~\eqref{eqn:three-factors} of a general matrix $\mtx{B}$, where the
outer factors $\mtx{V} \in \{0, 1\}^{n \times r}$
and $\mtx{W} \in \{0, 1\}^{m \times r}$ take binary values.
They developed an algorithm that
gives a rigorous tradeoff between the number $r$
of terms in the cut decomposition and the approximation error
as measured in cut norm.

Subsequently, Alon and Naor~\cite{AN06:Cut-Norm} developed
another algorithm for the cut decomposition
that proceeds by a rigorous process of iterative deflation. 
More precisely, Alon and Naor explain how to use semidefinite relaxation
and rounding to obtain a pair $(\vct{x}, \vct{y})$ where
the cut norm~\eqref{eqn:binary-opt} is approximately achieved.
They update the matrix $\mtx{B}$
via the rule $\mtx{B} \mapsto \mtx{B} - \lambda \vct{xy}^\adj$ where
$\lambda = \vct{x}^\adj \mtx{B} \vct{y} / (\norm{\vct{x}}_1 \norm{\smash{\vct{y}}}_1)$.
Iterating this approach leads to a tradeoff between the number $r$ of terms
in the cut decomposition and the cut norm approximation error
that improves substantially over~\cite{FK99:Cut-Decomposition}.

\subsubsection{Sign and binary component decompositions}

Our work introduces exact semidiscrete factorizations,
where the left factor $\mtx{V}$ consists of signs $\{\pm 1\}$
or binary values $\{0, 1\}$.  In this paper, we consider
the positive-semidefinite case, where the left and right
factors match: $\mtx{W} = \mtx{V}$.  In the companion
work, we consider the asymmetric case where the right factor
$\mtx{W}$ is arbitrary (but might be discrete).
In particular, we see that the binary component
decomposition~\eqref{eq:01_factor}
is an exact symmetric cut decomposition.

Our work gives conditions for existence, uniqueness,
and computational tractability of these factorizations.
A serious limitation, however, is the restriction to
matrices that have low rank.  We will seek extensions
to general matrices in our ongoing research.

\subsection{Other approaches to structured factorization}

Because of its importance in data analysis,
there is an extensive literature on structured matrix factorization.
Some of the most popular examples of constrained factorization are
independent component analysis~\cite{Com94:Independent-Component},
nonnegative matrix approximation~\cite{PT94:Positive-Matrix},
sparse coding~\cite{OF96:Emergence-Simple-Cell},
and sparse principal component analysis~\cite{ZHT06:Sparse-PCA}.
Some frameworks for thinking about structured matrix factorization
appear in~\cite{TB99:Probabilistic-Principal,CDS02:Generalization-Principal,
Tro04:Phd-Thesis,Sre04:Learning-Matrix,
Wit10:Phd-Thesis,Jag11:Phd-Thesis,
Bac13:Convex-Relaxations,Ude15:Phd-Thesis,BE16:Hierarchical-Compound-Poisson-Factorization,
Bru17:Phd-Thesis,HV19:Low-Rank},
among numerous other sources.  In this section,
we summarize a few other methods that have been
proposed for matrix factorization.
See~\cite{Ude15:Phd-Thesis,Bru17:Phd-Thesis}
for more complete literature reviews.

\subsubsection{Deflation}

A large number of authors have proposed matrix factorization
techniques based on
deflation~\cite{Wit10:Phd-Thesis,Jag11:Phd-Thesis,
Bac13:Convex-Relaxations,Ude15:Phd-Thesis,Bru17:Phd-Thesis}.
The basic step in these methods
(attempts) to solve a problem like
\begin{equation} \label{eqn:pmd}
\underset{\vct{x} \in \R^{n}, \vct{y}\in \R^{m}}{\maximize}\quad
	\vct{x}^\adj \mtx{B} \vct{y}
	\quad\subjto\quad
	P_1(\vct{x}) \leq c_1
	\quad\text{and}\quad
	P_2(\vct{y}) \leq c_2.
\end{equation}
The functions $P_1$ and $P_2$ are (convex) regularizers that promote
structure in the rank-one factor $\vct{xy}^\adj$.  For example,
when $P_1$ is the $\ell_1$ norm, this formulation tends to promote
sparsity in the factors.
Given an approximate solution $(\vct{x}, \vct{y})$,
we update the matrix via~\eqref{eqn:deflation-classic}
or using the conditional gradient method~\cite{Jag11:Phd-Thesis}.
Other deflation techniques have been developed for the particular task of sparse principal component analysis \cite{Mac09:Deflation}.
Witten~\cite{Wit10:Phd-Thesis} refers to this approach
as \emph{penalized matrix decomposition}.

In rare cases where we can provably compute an approximate
solution to~\eqref{eqn:pmd}, this approach leads to a
rigorous tradeoff between the number of terms in the
matrix approximation and the Frobenius-norm approximation
error~\cite{Jag11:Phd-Thesis}.  Unfortunately,
the core optimization~\eqref{eqn:pmd} is usually
computationally hard, and all bets are off.

In practice, the most common heuristic
for~\eqref{eqn:pmd} is alternating maximization.
Other nonconvex optimization algorithms,
such as projected gradient, can also be applied and often yield better results \cite{Ude15:Phd-Thesis}.
In some special cases, we can attack the problem~\eqref{eqn:pmd}
via semidefinite relaxation~\cite{AEJ07:Sparse-PCA}.
To the best of our knowledge, none of these methods
offer any guarantees.

\subsubsection{Direct methods}

Many other authors~\cite{Gor03:Generalized2-Linear2,Tro04:Phd-Thesis,Sre04:Learning-Matrix,
Bac13:Convex-Relaxations,Ude15:Phd-Thesis,Bru17:Phd-Thesis,HV19:Low-Rank}
have considered problem formulations
that seek to compute the matrix factors directly.
These approaches frame an optimization problem like
\begin{equation} \label{eqn:matrix-approx}
\underset{\mtx{V} \in \R^{n \times r}, \mtx{W} \in \R^{m \times r}}{\minimize}\quad
	\ell(\mtx{B}, \mtx{VW}^\adj) + P_1(\mtx{V}) + P_2(\mtx{W}).
\end{equation}
The loss function $\ell(\cdot, \cdot)$ is typically the Frobenius norm,
and the functions $P_1$ and $P_2$ are (convex) penalties that promote
structure on the matrix factors.

In practice, the most common heuristic for solving~\eqref{eqn:matrix-approx}
is alternating minimization.  Other nonconvex algorithms, such
as projected gradient, can also be applied.  When the inner dimension
$r$ of the factors is large enough, the optimization
problem~\eqref{eqn:matrix-approx} is sometimes tractable,
in spite of the apparent nonconvexity;
see~\cite{Gor03:Generalized2-Linear2,Bac13:Convex-Relaxations,Bru17:Phd-Thesis,HV19:Low-Rank}.
Nevertheless, we must regard~\eqref{eqn:matrix-approx}
as a heuristic, except in a limited set of circumstances.

\subsection{Conclusions}

The results presented here take the reverse point of view from most of the existing literature.
We first identify a class of matrices that admits a unique discrete factorization,
and we use this insight to develop a tractable algorithm that provably computes this factorization.
The next step is to understand how to find an approximate discrete factorization of a noisy matrix.
We believe that this kind of factorization has several applications, including
activity detection in large asynchronous networks.

\appendix

\section{Dimension reduction for Algorithm~\ref{alg:symfactor}} \label{sub:dimension_reduction}

As it is stated, Algorithm~\ref{alg:symfactor} requires solving semidefinite programs
in an $n \times n$ matrix variable.  In this section, we develop an equivalent
procedure that optimizes over a much lower-dimensional space of matrices.
This approach, which we document in Algorithm~\ref{alg:symfactor-redux},
has significantly lower resource usage.

\begin{theorem}[Efficient sign component decomposition] \label{thm:correct-redux}
Let $\mtx{A} \in \Sym_n$ be a rank-r correlation matrix that admits
a sign component decomposition
\begin{displaymath}
\mtx{A} = \sum_{i=1}^r \tau_i \vct{s}_i \vct{s}_i^\transp
\quad\text{where}\quad
\text{$\vct{s}_i \in \{ \pm 1 \}^n$ and $(\tau_1, \dots, \tau_r) \in \Delta_r^+$.}
\end{displaymath}
Assume that the family $\mathcal{S} = \{ \vct{s}_1, \dots, \vct{s}_r \}$
of sign components is Schur independent.  Then
Algorithm~\ref{alg:symfactor-redux} computes the
sign component decomposition up to trivial symmetries.
That is, the output is the unordered set of pairs
$ \{ (\tau_i, \xi_i \vct{s}_i ) : 1 \leq i \leq r \}$,
where $\xi_i \in \{ \pm 1 \}$ are signs.
This algorithm can be implemented
with arithmetic cost $\mathcal{O}(n^3 \mathrm{polylog}(r))$.
\end{theorem}

Up to logarithmic factors in $r$, the running time for Algorithm~\ref{alg:symfactor-redux}
matches the cost of computing a full eigenvalue decomposition of a dense $n \times n$ symmetric matrix.

\begin{proof}
Let $\mathcal{F} = \conv \{ \vct{ss}^\transp : \vct{s} \in \mathcal{S} \} \subset \mathcal{E}_n$
be the simplicial face of the elliptope that contains the matrix $\mtx{A}$.
Let $\mtx{Q}=\mathrm{orth}(\mtx{A}) \in \R^{n \times r}$ be a matrix with orthonormal columns that span the range of $\mtx{A}$.
In particular, $\mtx{P} = \mtx{QQ}^\transp$ is the orthogonal projector onto the range of $\mtx{A}$.
We can use these matrices to compress all of the optimization problems
that arise in Algorithm~\ref{alg:symfactor}.

We begin with the random optimization problem~\eqref{eqn:random-opt-pf}.
It is not hard to check that the feasible set of~\eqref{eqn:random-opt-pf}
can be rewritten as follows.  Let $\vct{q}_j^\transp \in \R^{r}$ be the $i$th row
of the matrix $\mtx{Q}$.  Then
\begin{displaymath}
\mathcal{F} = \{ \mtx{X} : \text{$\trace(\mtx{PX}) = n$ and $\mtx{X} \in \mathcal{E}_n$} \}
	= \{ \mtx{QYQ}^\transp : \text{$\vct{q}_j^\transp \mtx{Y} \vct{q}_j = 1$ for each $j$
	and $\mtx{Y} \psdge \mtx{0}$} \}.
\end{displaymath}
Indeed, recall that $\mtx{P}\vct{s} = \vct{s}$ for $\vct{s} \in \mathcal{S}$.
Each feasible point $\mtx{X}$ for~\eqref{eqn:random-opt-pf} belongs
to the face $\mathcal{F}$, so it must satisfy $\mtx{X} = \mtx{PXP}$.  Expanding
the orthogonal projectors, we obtain the parameterization $\mtx{X} = \mtx{QYQ}^\transp$
where $\mtx{Y} = \mtx{Q}^\transp \mtx{X} \mtx{Q} \in \Sym_r$.  
Moreover, $\mtx{X}$ is psd. According to the conjugation rule (Fact~\ref{fact:psd_invariance}),
this is equivalent to demanding $\mtx{Y} \psdge \mtx{0}$
Finally, the diagonal contraints $\mathbf{e}_j^\transp \mtx{X} \mathbf{e}_j = 1$
translate directly into the conditions $\vct{q}_j^\transp \mtx{Y} \vct{q}_j = 1$
for each index $j$.

We can conjugate the last display by the orthonormal matrix $\mtx{Q}$ to see that
\begin{displaymath}
\tilde{\mathcal{F}} = \conv\{ \mtx{Q}^\transp \vct{s}\vct{s}^\transp \mtx{Q} : \vct{s} \in \mathcal{S} \}
	= \{ \mtx{Y} \in \Sym_r : \text{$\vct{q}_j^\transp \mtx{Y} \vct{q}_j = 1$ for each $j$
	and $\mtx{Y} \psdge \mtx{0}$} \}.
\end{displaymath}
Moreover, the set $\tilde{\mathcal{F}}$ on the left-hand side is a simplex.
As a consequence, we can draw a standard normal vector $\vct{g} \in \R^r$ and
solve the optimization problem
\begin{equation} \label{eqn:random-opt-sm}
\underset{\mtx{Y} \in \Sym_r}{\maximize} \quad \vct{g}^\transp \mtx{Y} \vct{g}
	\quad\subjto\quad \text{$\vct{q}_j^\transp \mtx{Y} \vct{q}_j = 1$ for each $j$
	and $\mtx{Y} \psdge \mtx{0}$}.
\end{equation}
According to Lemma~\ref{lem:random}, the unique solution will be a matrix
$\mtx{Y}_{\star} = \mtx{Q}^\transp \vct{ss}^\transp \mtx{Q}$
for some $\vct{s} \in \mathcal{S}$.

The deflation step of Algorithm~\ref{alg:symfactor}
can also be mapped down to the simplex $\tilde{\mathcal{F}}$.
We just need to solve
\begin{displaymath}
\underset{\zeta \in \R}{\maximize} \quad \zeta
\quad\subjto\quad
\zeta (\mtx{Q}^\transp \mtx{A} \mtx{Q}) + (1 - \zeta) \mtx{Y}_{\star} \psdge \mtx{0}. 
\end{displaymath}
As before, Lemma~\ref{lem:deflation} ensures that this procedure extracts the
rank-one component $\mtx{Y}_{\star}$ from $\mtx{Q}^\transp \mtx{A} \mtx{Q}$,
decreasing the rank by one.

Next, let us propose a further simplification to the random optimization
problem~\eqref{eqn:random-opt-sm} by noticing that the equality constraints
are always redundant.  First, rewrite the constraints as
\begin{displaymath}
\trace( \vct{q}_j \vct{q}_j^\transp \mtx{Y} ) = 1
\quad\text{for $j = 1, \dots, n$.}
\end{displaymath}
Of course, each constraint matrix $\vct{q}_j \vct{q}_j^\transp \in \Sym_r$.
In fact, the constraint matrices also satisfy additional affine constraints.
For each $\vct{s} \in \mathcal{S}$, we calculate that
\begin{displaymath}
\trace( \vct{q}_j \vct{q}_j^\transp \mtx{Q}^\transp \vct{ss}^\transp \mtx{Q} )
	= \trace( \mtx{Q}^\transp \mathbf{e}_j \mathbf{e}_j^\transp \mtx{Q}\mtx{Q}^\transp \vct{ss}^\transp \mtx{Q} )
	= \trace( \mathbf{e}_j \mathbf{e}_j^\transp \mtx{P}\vct{ss}^\transp \mtx{P})
	= \trace( \mathbf{e}_j \mathbf{e}_j^\transp \vct{ss}^\transp)
	= \ip{ \mathbf{e}_j }{ \vct{s} }^2 = 1.
\end{displaymath}
Therefore, the constraint matrices lie in an affine subspace of $\Sym_r$
with dimension $\binom{r+1}{2} - r  = \binom{r}{2} + 1$.
We can  select a maximal linearly independent subset of
$\{ \vct{q}_j \vct{q}_j^\transp : j = 1, \dots, n \}$
and enforce this smaller family of constraints. 

To conclude, observe that Algorithm~\ref{alg:symfactor-redux} involves
$2(r-1)$ semidefinite programs with variable $\mtx{X} \in \Sym_r$.
The number of affine constraints in each SDP is bounded by $\binom{r}{2}+2$.
Standard interior point solvers~\cite{AHO98:Primal-Dual-Interior-Point}
can solve such problems to fixed accuracy in time $\mathcal{O}(r^{6.5})$.
This bound can ostensibly be improved to $\mathcal{O}(r^5 \mathrm{polylog}(r))$
using a method proposed in the theoretical algorithms
literature~\cite[Table 2]{LSW15:Convex-Optimization}.

Finally, we must account for the cost of computing a basis for the
range of the input matrix $\mtx{A}$ and lifting the solution from
the lower-dimensional space back to the original sign components.
For Algorithm~\ref{alg:symfactor-redux},
this leads to a total runtime of $\mathcal{O}(n^3 \mathrm{polylog}(r) )$
using the theoretical method.  This bound relies on the
restriction~\eqref{eq:factorization_rank}
that $r = \mathcal{O}(\sqrt{n})$  for Schur independence.
\end{proof}

\begin{algorithm}[t]
\algnewcommand{\Input}[1]{%
  \State \textbf{Input:}
  \Statex \hspace*{\algorithmicindent}\parbox[t]{.8\linewidth}{\raggedright #1}
}
\algnewcommand{\Output}[1]{%
  \State \textbf{Output:}
  \Statex \hspace*{\algorithmicindent}\parbox[t]{.8\linewidth}{\raggedright #1}
}
\algnewcommand{\Initialize}[1]{%
  \State \textbf{Initialize:}
  \Statex \hspace*{\algorithmicindent}\parbox[t]{.8\linewidth}{\raggedright #1}
}
{\small 
\begin{algorithmic}[1]
\caption{{\small \textit{Efficient sign component decomposition~\eqref{eq:binary_factorization}
of a matrix with Schur independent components}. 
Implements the procedure in Theorem~\ref{thm:correct-redux}.}}
\label{alg:symfactor-redux}

\Require	Rank-$r$ correlation matrix $\mtx{A} \in \Sym_n$ that satisfies~\eqref{eqn:A-for-alg}
\Ensure		Sign components $\{\tilde{\vct{s}}_1,\dots,\tilde{\vct{s}}_r\} \subset \{\pm 1\}^n$ and convex coefficients $\tilde{\vct{\tau}} \in \Delta_r^+$ where $\mtx{A} = \sum_{i=1}^r \tilde{\tau}_i \tilde{\vct{s}}_i \tilde{\vct{s}}_i^\transp$
\Statex
\Function{EfficientSignComponentDecomposition}{$\mtx{A}$}

\State	$[n, \sim] \gets \texttt{size}(\mtx{A})$ and $r \gets \rank(\mtx{A})$

\State	$\mtx{Q} \gets \texttt{orth}(\mtx{A})$
	\Comment Orthonormal basis for the range of $\mtx{A}$
\State	$\mtx{M} \gets \mtx{Q}^\transp \mtx{A} \mtx{Q}$
	\Comment Compress the input matrix
\State	Use \textsf{RRQR} to find a maximal independent set of constraints:
\begin{displaymath}
\underset{J \subseteq \{1,\dots,n\}}{\maximize}\quad \abs{J}
\quad\subjto\quad
\text{$\{ \vct{q}_j \vct{q}_j^\transp : j \in J \}$ is linearly independent}
\end{displaymath}
\For{$i = 1$ to $(r-1)$}
\State	$\vct{g} \gets \texttt{randn}(r, 1)$
	\Comment Draw a random direction
\State	Find the solution $\mtx{Y}_{\star}$ to the semidefinite program
\Comment	Step 1
\begin{displaymath}
\underset{\mtx{Y} \in \Sym_r}{\maximize} \quad \vct{g}^\transp \mtx{Y} \vct{g}
\quad\subjto\quad \text{$\vct{q}_j^\transp \mtx{Y} \vct{q}_j = 1$ for $j \in J$ and $\mtx{Y} \psdge \mtx{0}$}
\end{displaymath}
\State	Factorize the rank-one matrix $\mtx{Y}_{\star} = \mtx{Q}^\transp\tilde{\vct{s}}_i \tilde{\vct{s}}_i^\transp \mtx{Q}$
\Comment	Extract a sign component 
\State	Find the solution $\zeta_{\star}$ to the semidefinite program 
\Comment	Step 2
\begin{displaymath}
\underset{\zeta \in \R}{\maximize} \quad \zeta \quad \subjto \quad
\zeta \mtx{M} + (1-\zeta) \mtx{Y}_{\star} \psdge \mtx{0}
\end{displaymath}
\State	$\mtx{M} \gets \zeta_{\star} \mtx{M} + (1-\zeta_{\star}) \mtx{Y}_{\star}$
\Comment	Step 3
\EndFor
\State	Factorize the rank-one matrix $\mtx{M} = \mtx{Q}^\transp\tilde{\vct{s}}_r\tilde{\vct{s}}_r^\transp\mtx{Q}$
\Comment	$\rank(\mtx{M}) = 1$ in final iteration
\State	Find the unique solution $\tilde{\vct{\tau}} \in \R^r$ to the linear system
\Comment	Step 4
\begin{displaymath}
\mtx{M} = \sum_{i=1}^r \tilde{\tau}_i \mtx{Q}^\transp \tilde{\vct{s}}_i \tilde{\vct{s}}_i^\transp \mtx{Q}
\end{displaymath}
\EndFunction
\end{algorithmic}
}
\end{algorithm}

\section*{Acknowledgments}

The authors thank Benjamin Recht for helpful conversations at an early stage of this project and Madeleine Udell for valuable comments regarding the related work section.
Peter Jung suggested activity detection in massive MIMO system as a potential application.
This research was partially funded by ONR awards N00014-11-1002, N00014-17-12146, and
N00014-18-12363.  Additional support was provided by the Gordon \& Betty Moore Foundation.

\bibliographystyle{myalpha}
\bibliography{binfactor}

\begin{thebibliography}{dEGJL07}

\bibitem[AHO98]{AHO98:Primal-Dual-Interior-Point}
F.~Alizadeh, J.-P.~A. Haeberly, and M.~L. Overton.
\newblock Primal-dual interior-point methods for semidefinite programming:
  convergence rates, stability and numerical results.
\newblock {\em SIAM J. Optim.}, 8(3):746--768, 1998.

\bibitem[AN06]{AN06:Cut-Norm}
N.~Alon and A.~Naor.
\newblock Approximating the cut-norm via {G}rothendieck's inequality.
\newblock {\em SIAM J. Comput.}, 35(4):787--803, 2006.

\bibitem[Bac13]{Bac13:Convex-Relaxations}
F.~R. Bach.
\newblock Convex relaxations of structured matrix factorizations.
\newblock Available at \url{http://arXiv.org/abs/1309.3117}, 2013.

\bibitem[Bar02]{Bar02:Course-Convexity}
A.~Barvinok.
\newblock {\em A course in convexity}, volume~54 of {\em Graduate Studies in
  Mathematics}.
\newblock American Mathematical Society, Providence, RI, 2002.

\bibitem[BE16]{BE16:Hierarchical-Compound-Poisson-Factorization}
M.~Basbug and B.~Engelhardt.
\newblock Hierarchical compound poisson factorization.
\newblock In M.~F. Balcan and K.~Q. Weinberger, editors, {\em Proceedings of
  The 33rd International Conference on Machine Learning}, volume~48 of {\em
  Proceedings of Machine Learning Research}, pages 1795--1803, New York, New
  York, USA, 20--22 Jun 2016. PMLR.

\bibitem[Bha97]{Bha97:Matrix-Analysis}
R.~Bhatia.
\newblock {\em Matrix analysis}, volume 169 of {\em Graduate Texts in
  Mathematics}.
\newblock Springer-Verlag, New York, 1997.

\bibitem[BLX18]{BLX18:Multivariate-Approximation}
A.~D. Barbour, M.~J. Luczak, and A.~Xia.
\newblock Multivariate approximation in total variation, {I}: {E}quilibrium
  distributions of {M}arkov jump processes.
\newblock {\em Ann. Probab.}, 46(3):1351--1404, 2018.

\bibitem[Bru17]{Bru17:Phd-Thesis}
J.~J. Bruer.
\newblock {\em Recovering structured low-rank operators using nuclear norms}.
\newblock PhD thesis, Caltech, Pasadena, 2017.

\bibitem[CDS02]{CDS02:Generalization-Principal}
M.~Collins, S.~Dasgupta, and R.~E. Schapire.
\newblock A generalization of principal component analysis to the exponential
  family.
\newblock In {\em Adv. Neural Information Processing Systems}, 2002.

\bibitem[Com94]{Com94:Independent-Component}
P.~Comon.
\newblock Independent component analysis, a new concept?
\newblock {\em Signal Process}, 36(3):287 -- 314, 1994.

\bibitem[CSY18]{CSY18:MIMO-AMP}
Z.~{Chen}, F.~{Sohrabi}, and W.~{Yu}.
\newblock Sparse activity detection for massive connectivity.
\newblock {\em IEEE Transactions on Signal Processing}, 66(7):1890--1904, April
  2018.

\bibitem[dEGJL07]{AEJ07:Sparse-PCA}
A.~d'Aspremont, L.~El~Ghaoui, M.~I. Jordan, and G.~R.~G. Lanckriet.
\newblock A direct formulation for sparse {PCA} using semidefinite programming.
\newblock {\em SIAM Rev.}, 49(3):434--448, 2007.

\bibitem[DL97]{DL97:Geometry-Cuts}
M.~M. Deza and M.~Laurent.
\newblock {\em Geometry of cuts and metrics}, volume~15 of {\em Algorithms and
  Combinatorics}.
\newblock Springer-Verlag, Berlin, 1997.

\bibitem[FJ19]{FJ19:Khatri-Rao-RIP}
A.~Fengler and P.~Jung.
\newblock On the restricted isometry property of centered self {Khatri-Rao}
  products.
\newblock {\em arXiv preprint arXiv:1905.09245}, 2019.

\bibitem[FK99]{FK99:Cut-Decomposition}
A.~Frieze and R.~Kannan.
\newblock Quick approximation to matrices and applications.
\newblock {\em Combinatorica}, 19(2):175--220, 1999.

\bibitem[GLS93]{GLS93:Combinatorial-Optimization}
M.~Gr\"{o}tschel, L.~Lov\'{a}sz, and A.~Schrijver.
\newblock {\em Geometric algorithms and combinatorial optimization}, volume~2
  of {\em Algorithms and Combinatorics}.
\newblock Springer-Verlag, Berlin, second edition, 1993.

\bibitem[Gor03]{Gor03:Generalized2-Linear2}
G.~J. Gordon.
\newblock Generalized{$^2$} linear{$^2$} models.
\newblock In S.~Becker, S.~Thrun, and K.~Obermayer, editors, {\em Advances in
  Neural Information Processing Systems 15}, pages 593--600. MIT Press, 2003.

\bibitem[Gru07]{Gru07:Convex-Geometry}
P.~M. Gruber.
\newblock {\em Convex and discrete geometry}, volume 336 of {\em Grundlehren
  der Mathematischen Wissenschaften [Fundamental Principles of Mathematical
  Sciences]}.
\newblock Springer, Berlin, 2007.

\bibitem[Her51]{Her51:Normal-Form}
C.~Hermite.
\newblock Sur l'introduction des variables continues dans la th\'{e}orie des
  nombres.
\newblock {\em J. Reine Angew. Math.}, 41:191--216, 1851.

\bibitem[HJC18]{HJC18:Activity-Detection}
S.~Haghighatshoar, P.~Jung, and G.~Caire.
\newblock Improved scaling law for activity detection in massive mimo systems.
\newblock In {\em 2018 IEEE International Symposium on Information Theory
  (ISIT)}, pages 381--385, June 2018.

\bibitem[HUL01]{HL01:Fundamentals-Convex}
J.-B. Hiriart-Urruty and C.~Lemar\'{e}chal.
\newblock {\em Fundamentals of convex analysis}.
\newblock Grundlehren Text Editions. Springer-Verlag, Berlin, 2001.
\newblock Abridged version of {{\i}t Convex analysis and minimization
  algorithms. I} [Springer, Berlin, 1993; MR1261420 (95m:90001)] and {{\i}t II}
  [ibid.; MR1295240 (95m:90002)].

\bibitem[HV19]{HV19:Low-Rank}
B.~D. {Haeffele} and R.~{Vidal}.
\newblock Structured low-rank matrix factorization: Global optimality,
  algorithms, and applications.
\newblock {\em IEEE T Pattern Anal}, pages 1--1, 2019.

\bibitem[Jag11]{Jag11:Phd-Thesis}
M.~Jaggi.
\newblock {\em Sparse convex optimization methods for machine learning}.
\newblock PhD thesis, ETH Z{\"u}rich, 2011.

\bibitem[Jol02]{Jol02:Principal-Component}
I.~T. Jolliffe.
\newblock {\em Principal component analysis}.
\newblock Springer Series in Statistics. Springer-Verlag, New York, second
  edition, 2002.

\bibitem[Kar72]{Kar72:21-NP-Problems}
R.~M. Karp.
\newblock Reducibility among combinatorial problems.
\newblock In {\em Complexity of computer computations ({P}roc. {S}ympos., {IBM}
  {T}homas {J}. {W}atson {R}es. {C}enter, {Y}orktown {H}eights, {N}.{Y}.,
  1972)}, pages 85--103. Plenum, New York, 1972.

\bibitem[KB79]{KB79:Polynomial-Algorithms}
R.~Kannan and A.~Bachem.
\newblock Polynomial algorithms for computing the {S}mith and {H}ermite normal
  forms of an integer matrix.
\newblock {\em SIAM J. Comput.}, 8(4):499--507, 1979.

\bibitem[Kol98]{Kol98:Phd-Thesis}
T.~G. Kolda.
\newblock {\em Limited-memory matrix methods with applications}.
\newblock PhD thesis, University of Michigan, 1998.

\bibitem[KT19]{KT19:Binary-Factorization-II}
R.~Kueng and J.~Tropp.
\newblock Binary component decomposition {Part II:} {T}he asymmetric case.
\newblock {\em preprint}, 2019.

\bibitem[LP96]{LP96:Facial-Structure}
M.~Laurent and S.~Poljak.
\newblock On the facial structure of the set of correlation matrices.
\newblock {\em SIAM J. Matrix Anal. Appl.}, 17(3):530--547, 1996.

\bibitem[LSW15]{LSW15:Convex-Optimization}
Y.~T. Lee, A.~Sidford, and S.~C.~W. Wong.
\newblock A faster cutting plane method and its implications for combinatorial
  and convex optimization.
\newblock In {\em 2015 IEEE 56th Annual Symposium on Foundations of Computer
  Science}, pages 1049--1065, Oct 2015.

\bibitem[LY18]{LY18:MIMO}
L.~Liu and W.~Yu.
\newblock Massive connectivity with massive {MIMO}---{P}art {I}: {D}evice
  activity detection and channel estimation.
\newblock {\em IEEE Trans. Signal Process.}, 66(11):2933--2946, 2018.

\bibitem[Mac09]{Mac09:Deflation}
L.~W. Mackey.
\newblock Deflation methods for sparse pca.
\newblock In D.~Koller, D.~Schuurmans, Y.~Bengio, and L.~Bottou, editors, {\em
  Advances in Neural Information Processing Systems 21}, pages 1017--1024.
  Curran Associates, Inc., 2009.

\bibitem[OF96]{OF96:Emergence-Simple-Cell}
B.~A. Olshausen and D.~J. Field.
\newblock Emergence of simple-cell receptive field properties by learning a
  sparse code for natural images.
\newblock {\em Nature}, 381(6583):607--609, 1996.

\bibitem[OP83]{LP83:Digital-Compression}
D.~O'Leary and S.~Peleg.
\newblock Digital image compression by outer product expansion.
\newblock {\em IEEE T. Commun.}, 31(3):441--444, March 1983.

\bibitem[PT94]{PT94:Positive-Matrix}
P.~Paatero and U.~Tapper.
\newblock Positive matrix factorization: A non-negative factor model with
  optimal utilization of error estimates of data values.
\newblock {\em Environmetrics}, 5(2):111--126, 1994.

\bibitem[Roc70]{Roc70:Convex-Analysis}
R.~T. Rockafellar.
\newblock {\em Convex analysis}.
\newblock Princeton Mathematical Series, No. 28. Princeton University Press,
  Princeton, N.J., 1970.

\bibitem[Sch14]{Sch13:Convex-Bodies}
R.~Schneider.
\newblock {\em Convex bodies: the {B}runn-{M}inkowski theory}, volume 151 of
  {\em Encyclopedia of Mathematics and its Applications}.
\newblock Cambridge University Press, Cambridge, expanded edition, 2014.

\bibitem[Smi61]{Smi61:Smith-Normal}
H.~J.~S. Smith.
\newblock On systems of linear indeterminate equations and congruences.
\newblock {\em Philos. Tr. Soc. Lon.}, 151:293--326, 1861.

\bibitem[Sre04]{Sre04:Learning-Matrix}
N.~Srebro.
\newblock {\em Learning with matrix factorizations}.
\newblock ProQuest LLC, Ann Arbor, MI, 2004.
\newblock Thesis (Ph.D.)--Massachusetts Institute of Technology.

\bibitem[TB99]{TB99:Probabilistic-Principal}
M.~E. Tipping and C.~M. Bishop.
\newblock Probabilistic principal component analysis.
\newblock {\em J. R. Stat. Soc. Ser. B Stat. Methodol.}, 61(3):611--622, 1999.

\bibitem[Tro04]{Tro04:Phd-Thesis}
J.~A. Tropp.
\newblock {\em Topics in sparse approximation}.
\newblock PhD thesis, The University of Texas at Austin, 2004.

\bibitem[Tro12]{Tro12:Tail-Bounds}
J.~A. Tropp.
\newblock User-friendly tail bounds for sums of random matrices.
\newblock {\em Found. Comput. Math.}, 12(4):389--434, 2012.

\bibitem[Tro15]{Tro15:Integer-Factorization}
J.~A. Tropp.
\newblock Integer factorization of a positive-definite matrix.
\newblock {\em SIAM J. Discrete Math.}, 29(4):1783--1791, 2015.

\bibitem[Tro18]{Tro18:Simplicial-Faces}
J.~A. Tropp.
\newblock Simplicial faces of the set of correlation matrices.
\newblock {\em Discrete Comput. Geom.}, 60(2):512--529, 2018.

\bibitem[Ude15]{Ude15:Phd-Thesis}
M.~Udell.
\newblock {\em Generalized low rank models}.
\newblock PhD thesis, Stanford University, 2015.

\bibitem[Wit10]{Wit10:Phd-Thesis}
D.~M. Witten.
\newblock {\em A penalized matrix decomposition and its applications}.
\newblock PhD thesis, Stanford University, 2010.

\bibitem[Yap00]{Yap00:Fundamental-Problems}
C.~K. Yap.
\newblock {\em Fundamental problems of algorithmic algebra}.
\newblock Oxford University Press, New York, 2000.

\bibitem[ZHT06]{ZHT06:Sparse-PCA}
H.~Zou, T.~Hastie, and R.~Tibshirani.
\newblock Sparse principal component analysis.
\newblock {\em J. Comput. Graph. Statist.}, 15(2):265--286, 2006.

\end{thebibliography}

\end{document}